\documentclass[journal]{IEEEtran}
\IEEEoverridecommandlockouts
\usepackage{cite}
\usepackage{amsmath,amssymb,amsfonts}
\usepackage{algorithm}
\usepackage{algpseudocode}
\usepackage{graphicx}
\usepackage{psfrag}
\usepackage{subcaption}

\graphicspath{{./images/}}
\DeclareGraphicsExtensions{.pdf,.jpeg,.png}
\usepackage{textcomp}
\usepackage{xcolor}

\newcommand{\bm}{\mathbf}
\newcommand{\be}{\begin{equation}}
\newcommand{\ee}{\end{equation}}
\newcommand{\bea}{\begin{eqnarray}}
\newcommand{\eea}{\end{eqnarray}}

\newcommand{\bG}{{\bf G}}

\begin{document}

\title{Joint Phase Noise and Channel Estimation for OTFS  \\
}

\author{\IEEEauthorblockN{Stephen McWade ~\IEEEmembership{Member, IEEE,}
 and 
Arman Farhang, ~\IEEEmembership{Senior Member, IEEE,}}

\thanks{
The authors are with the Department of Electronic and Electrical Engineering, Trinity College Dublin, Dublin 2, D02 PN40 (email: smcwade@tcd.ie, arman.farhang@tcd.ie). 

This publication has emanated from research conducted with the financial support of Research Ireland under the US-Ireland R{\&}D Partnership Programme Grant Numbers 24/US/4013 and 21/US/3757.}}

\maketitle

\begin{abstract}
This paper investigates the effect of oscillator phase noise in orthogonal time frequency space (OTFS) systems. The paper provides in-depth analysis of the interference due to phase noise in the delay-Doppler domain and derives expressions for SINR for three different oscillator types, namely free-running oscillators, continuous-time phase locked loops (PLLs) and discrete-time PLLs. The analysis demonstrates the OTFS is sensitive to phase noise and requires appropriate estimation and compensation. In particular, the analysis shows phase noise imposed inter-Doppler-interference (IDI) is severe and that existing phase noise estimation techniques which only consider the common-phase-error (CPE) can not compensate this IDI effectively. Additionally, the existing methods in the OTFS literature on phase noise assume the channel to be a known single tap channel. Hence, in this paper, we propose a method for joint channel and phase noise estimation using a Wiener filtering approach. Our proposed method exploits the statistical nature of both the phase noise and the Doppler spread channel. Our numerical results demonstrate the superior performance of our proposed technique, with gains of up to 8~dB in terms of bit error rate (BER) over existing methods in the literature.
\end{abstract}


\section{Introduction}
Orthogonal time frequency space (OTFS) modulation is a prominent modulation scheme for the next generation of wireless networks which has attracted a great deal of interest due to its robustness to time-varying channel effects \cite{hadani_2017, OTFS_hanzo}. OTFS harvests the full diversity of  the channel by spreading the data symbols across the whole time-frequency plane \cite{Farhang_letter}. Thus far, the majority of OTFS literature has been on the theoretical aspects of this waveform and there is only a small body of literature that considers practical aspects, such as RF impairments. In practical systems, RF impairments such as carrier frequency offset (CFO), in-phase (I) and quadrature-phase (Q) imbalances and oscillator phase noise are present \cite{RF_impairment_survey}. These RF impairments lead to significant performance degradation \cite{Marsalek_OTFS_RF}. Additionally, RF impairments become more severe at higher frequency bands which are being considered for use in the next generation of wireless networks \cite{Tataria_6G}. 

In this paper, we particularly focus on the effect of phase noise on OTFS, which has not been fully analyzed in the literature as of yet.  It is well-known that both phase noise and Doppler spread channels are time-varying processes \cite{Hajimari_PN_1998, MATZ20111}. However, the channel taps in Doppler spread channels are generally considered as low-pass processes while phase noise is a wideband process. This means that phase noise and Doppler spread are fundamentally different in nature and channel estimation methods which work for the latter may not work for the former. Additionally, since OTFS operates by taking data symbols at intervals spaced apart by $M$ and performing an inverse discrete Fourier transform (IDFT) operation on them, the sample spacing is $M$ times larger. Thus, the effective sample-to-sample variance of the phase noise is will be higher in the delay-Doppler domain than in an OFDM system which performs an IDFT operation on contiguous data samples. Therefore, OTFS systems may be more sensitive than OFDM to interference caused by phase noise. This highlights the importance of the study presented in this paper. 

 There is a large body of work on orthogonal frequency division multiplexing (OFDM) systems and it is well understood that phase noise can severely degrade the performance of OFDM systems due to the intercarrier interference (ICI) it introduces \cite{Fettweis_PN_OFDM, Armada_OFDM_PN_1998, Chung_OFDM_PN_2022}.  In contrast, OTFS in the presence of phase noise is still a nascent research topic. However, a number of papers on the topic have been published in recent years \cite{Surabhi_OTFS_PN_2019, Bello_OTFS_pn, Liang_OTFS_PN_2025}.
The authors of \cite{Surabhi_OTFS_PN_2019} are the first to study the effect of phase noise on OTFS. However, they did not perform any analysis to support this claim and they considered phase noise to be perfectly known and compensated at the receiver. This led to the conclusion that OTFS is more resilient than OFDM to phase noise. However, in practical systems, phase noise cannot be perfectly known. Without a thorough analysis, the sensitivity of OTFS to phase noise is still unknown. 

On the topic of phase noise estimation, the authors of \cite{Bello_OTFS_pn} present a method for common-phase-error (CPE) estimation by adopting the OFDM approach of using phase tracking reference pilots (PTRPs) in either the delay-time domain or the delay-Doppler domain. This CPE-only estimation approach is only applicable in low phase noise scenarios where phase noise can be assumed to be constant for a number of samples, i.e., over each delay block. However, this assumption is not realistic under moderate to high amounts of phase noise. Additionally, the authors in \cite{Bello_OTFS_pn} only consider a single tap, line-of-sight sub-THz channel model which is considered to be perfectly known. This is a limiting assumption as, in practical systems, it is not possible to separately estimate the channel and phase noise. More recently, the authors of \cite{Liang_OTFS_PN_2025} propose a method for estimating and compensating phase noise for OTFS. The method presented in \cite{Liang_OTFS_PN_2025} estimates the CPE in the time-frequency domain and returns to the delay-Doppler domain to compensate the phase noise. This type of CPE approach assumes that the phase noise variance is low enough such that the phase is approximately constant for each delay block in the OTFS grid and any residual interference is treated as noise. As we will show in our analysis presented later in this paper, OTFS is ,in-fact, quite sensitive to the interference caused by phase noise and the assumption of constant phase for each delay block is not appropriate, especially in high phase noise scenarios. Furthermore, none of the aforementioned literature provide in-depth analysis of phase noise and its effects on OTFS in the delay-Doppler domain.

In OTFS literature, early channel estimation work assumes the channel to be partially linear time-invariant (LTI) for blocks of $M$ samples, such as the threshold-based method presented in \cite{Raviteja_OTFS_2019}. However, this is not a realistic assumption in the presence of phase noise, which causes sample-to-sample fluctuations in the effective channel. More recent papers use interpolation based estimation methods to improve OTFS channel estimation in high mobility scenarios. In particular, the authors of \cite{Sanoop_PCP} and \cite{Thaj_OTSM_2021} used basis expansion model (BEM) and Spline interpolation methods, respectively, to improve channel estimation. BEM is well suited for low-pass processes such as Doppler spread channels as the variations of the channel are smooth sample-to-sample. However, phase noise is a wideband process and the variations are not smooth sample-to-sample. Therefore, BEM may not be appropriate for estimating phase noise in OTFS systems. For phase noise estimation, the authors of \cite{Bello_DFTsOFDM_PN_2023} present a method for sub-THz single carrier systems using Wiener interpolation. However, this work only considers a frequency flat line-of-sight channel without delay spread or Doppler spread. For an OTFS system affected by phase noise, channel estimation needs to consider both the delay and Doppler spread as well as the phase noise effect.

Despite these early works on this topic, there is yet to be an in-depth analytical study on the effects of phase noise in the delay-Doppler domain. Additionally, to the best of our knowledge, there is no work that considers joint estimation of phase noise and channel in the presence of both delay and Doppler spread. Hence, this paper addresses these gaps in the literature with the following contributions:

\begin{itemize}
    \item We provide a full analytical study of the effect of phase noise on OTFS while considering a doubly dispersive channel. We present the structure of the phase noise matrix and its underlying effect on delay-Doppler multiplexing. We also derive the input-output relationship in the delay-Doppler domain.
    \item We utilize the derived input-output relationship to obtain the signal-to-interference-plus-noise ratio (SINR) expression for OTFS in presence of phase noise. 
    \item We provide full derivations for multiple different types of oscillators, namely,  free-running oscillator, continuous-time phase locked loop (PLL) and discrete-time PLL. In particular, for the  free-running oscillator case, we obtain closed form expression for the SINR. Our analysis shows that OTFS is sensitive to phase noise without appropriate compensation. 
    \item Utilizing this analysis, we propose a  joint channel and phase noise estimation technique using  Wiener filtering to estimate the full effective channel. Our proposed method uses the statistical properties of both the phase noise and the Doppler spread channel based on the analytical results outlined above.
\end{itemize}

We analyze the effectiveness of the proposed phase noise and channel estimation technique and the validity of our mathematical derivations through simulations. We compare the performance our proposed technique to state-of-the-art methods in the literature under a range of different scenarios via numerical simulations. We evaluate the detection performance in terms of error vector magnitude (EVM) and bit error rate (BER) and demonstrate performance gains of up to 8~dB in terms BER over BEM based estimation. Additionally, we evaluate the channel estimation performance in terms of normalized mean squared error (NMSE) and demonstrate that our proposed technique provides improved performance compared to the existing methods in the literature. Additionally, we investigate the effect of bit encoding on performance and demonstrate that our proposed technique provides performance gains of up to 6~dB in terms of BER over the existing methods in coded systems.

The remainder of this paper is organized as follows. Section~II describes the system model for an OTFS system in the presence of phase noise. Section~III presents the interference analysis and SINR derivations for the system. Section~IV describes our proposed technique for joint phase noise and channel estimation. Section~V presents numerical results and discussions thereof. Finally, Section~VI concludes the paper.

 \subsubsection*{Notations} Superscripts ${(\cdot)^{\rm{T}}}$ and ${(\cdot)^{\rm{H}}}$ denote transpose and Hermitian transpose, respectively. Bold lower-case characters are used to denote vectors and bold upper-case characters are used to denote matrices. $x[n]$ denotes the $n$-th element of the vector $\mathbf{x}$. The function $\rm{vec}(\mathbf{X})$ vectorizes the matrix $\mathbf{X}$ by concatenating its columns to form a vector, and $\otimes$ represents the Kronecker product. The $p\times{p}$ identity matrix and $p \times q$ all-zero matrix are  denoted by $\mathbf{I}_p$ and $\mathbf{0}_{p\times{q}}$, respectively. Finally, $j = \sqrt{-1}$ represents the imaginary unit.

\section{System Model}
\begin{figure*}[t]
\centering
\includegraphics[scale=0.75]{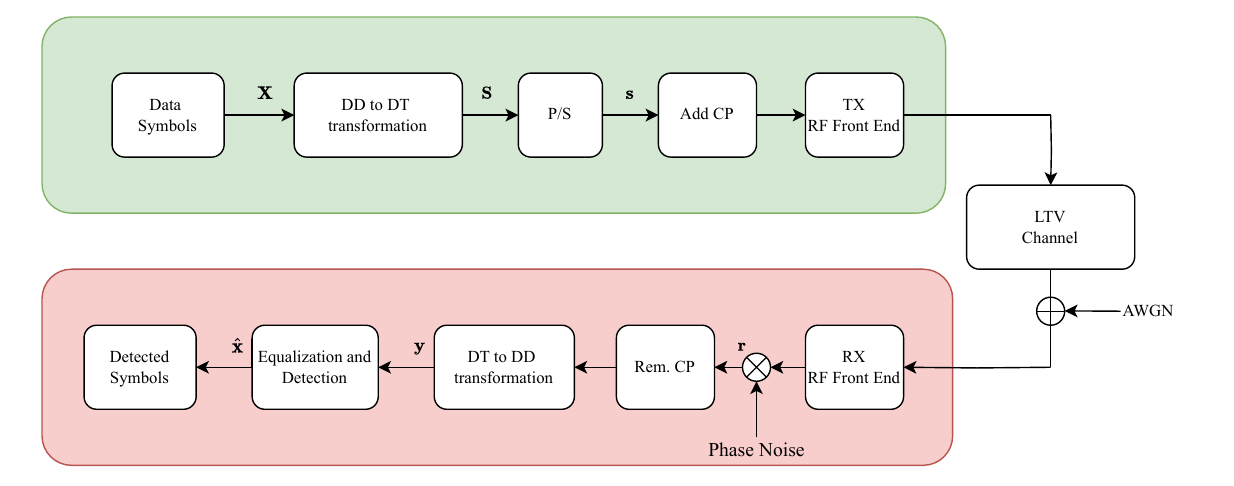}
\caption{Block diagram of the OTFS baseband equivalent transmitter and receiver.}\label{OTFS_blk_dia}
\end{figure*}
In this section, we present the OTFS system model in the presence of phase noise.  We consider an OTFS system with $M$ Doppler bins and $N$ delay bins with delay and Doppler spacings of $\Delta\tau$ and $\Delta\nu$, respectively \cite{RCP_OTFS_Rav}. Let the $M \times N$ matrix $\mathbf{X}$ contain the transmit quadrature amplitude modulation (QAM) data symbols on its elements in the delay-Doppler domain. The data symbols are assumed to be independent and identically distributed (i.i.d.) complex random variables of unit average power. In this work, similar to \cite{RCP_OTFS_Rav}, we consider OFDM-based OTFS modulation with rectangular pulse shape. Thus, the delay-time transmit signal is obtained by taking an $N$-point IDFT across the rows of $\mathbf{X}$, i.e., the Doppler dimension \cite{Farhang_letter}. Therefore, the delay-time domain transmit signal is given by
\begin{equation}
    \mathbf{S} = \mathbf{X}\mathbf{F}_{N}^{\rm{H}},
\end{equation}
where $\bm{F}_N$ is the $N$-point unitary discrete Fourier transform (DFT) matrix with $(l,k)$ elements $\frac{1}{\sqrt{N}}e^{-j\frac{2\pi}{N}lk}$ for $l,k=0,\ldots,N-1$. The signal then undergoes parallel to serial conversion and can be written in vectorized form as
\begin{equation}
    \mathbf{s} = \rm{vec}(\mathbf{S}) = (\mathbf{F}_{N}^{\rm{H}}\otimes\mathbf{I}_{M})\mathbf{x}.
\end{equation} 
where $\bm{x} = \rm{vec}(\bm{X})$. A cyclic prefix of length $N_{\rm{cp}}$ is appended to the beginning of the transmit signal. After analog-to-digital conversion and modulating the transmit signal to the carrier frequency,it propagates through the linear time-varying (LTV) channel. In practical systems, local oscillator imperfections induce unwanted phase noise into the signal at both the transmitter and receiver sides. However, when the phase noise bandwidth at the each side is small relative to the subcarrier spacing, the resulting phase noise effect is equivalent to a single phase noise process at the receiver with a bandwidth equal to the sum of the bandwidth of the transmitter and receiver side processes \cite{Fettweis_PN_OFDM}. Therefore, for ease of explanation, we consider a system where the signal is affected by phase noise at the receiver side only. Hence, the continuous-time received signal in baseband is represented as 
\begin{equation}
    r(t) = e^{j\theta(t)}\int \int h(\tau,\nu)s(t-\tau)e^{j2\pi\nu(t-\tau)}d\tau d\nu + \eta(t),
\end{equation}
where $h(\tau,\nu) = \sum_{p=0}^{P-1}h_{p}\delta(\tau - \tau_{p})\delta(\nu - \nu_{p}),$ is the delay-Doppler domain channel impulse response (CIR), which consists of $P$ channel paths, and $\eta(t)$ is the complex additive white Guassian noise (AWGN) with variance $\sigma_{\eta}^2$. The parameters $h_{p}$, $\tau_{p}$ and $\nu_{p}$ represent the channel gain, delay and Doppler shift associated with path $p$ of the channel,  respectively. The term $\theta(t)$ represents the continuous-time phase noise angle at time $t$.

\begin{figure*}[t]
    \normalsize
    \setcounter{equation}{9}
    \begin{equation}\mathbf{\Phi}_{\rm{DD}} = \left[
\begin{array}{cccccccccc}
&\phi_0[0] & 0 &\dots & 0 & \dots &\phi_1[0] & 0 &\dots & 0\\
& 0 &\phi_0[1] &{\dots} & 0& \dots & 0 &\phi_1[1] &{\dots} & 0 \\
&\boldsymbol{\vdots} &\boldsymbol{\vdots} &{\ddots} &\boldsymbol{\vdots}& \dots  &\boldsymbol{\vdots} &\boldsymbol{\vdots} &{\ddots} &\boldsymbol{\vdots}\\
& 0 & 0 &{\dots} &\phi_0[M-1]& \dots & 0 & 0 &{\dots} &\phi_1[M-1]\\
&\boldsymbol{\vdots} &\boldsymbol{\vdots} &{\ddots} &\boldsymbol{\vdots} & \dots &\boldsymbol{\vdots} &\boldsymbol{\vdots} &{\ddots} &\boldsymbol{\vdots}\\
&\phi_{N-1}[0] & 0 &\dots & 0& \dots &\phi_{0}[0] & 0 &\dots & 0\\
& 0 &\phi_{N-1}[1] &{\dots} & 0& \dots & 0 &\phi_{0}[1] &{\dots} & 0\\
&\boldsymbol{\vdots} &\boldsymbol{\vdots} &{\ddots} &\boldsymbol{\vdots} & \dots &\boldsymbol{\vdots} &\boldsymbol{\vdots} &{\ddots} &\boldsymbol{\vdots}\\
& 0 & 0 &{\dots} &\phi_{0}[M-1]& \dots & 0 & 0 &{\dots} &\phi_{0}[M-1]
\end{array}
\right] \label{eq10}\end{equation}
\end{figure*}
\setcounter{equation}{3}
The received signal for a given OTFS block is then sampled with sampling period $T_{\rm{s}}$ and the CP is removed. The discrete received signal samples can be expressed as 
 \begin{equation}
        r[n] =  e^{j\theta[n]}\sum_{l=0}^{L-1}h[n,l]s[n-l] + \eta[n], \label{eq:7}
\end{equation}
where $h[n,l]$ is the discrete-time baseband CIR for the channel at sample $n$ and delay $l$, $\theta[n]$ is the phase noise at sample $n$ and $L$ is the length of the discrete-time channel. By stacking the received signal samples into an $MN \times 1$ vector, (\ref{eq:7}) can be represented in matrix form as 
\begin{equation}
    \mathbf{r} = \mathbf{\Phi}_{\rm{DT}}\mathbf{H}_{\mathrm{DT}}\mathbf{s} + \boldsymbol{\eta},
\end{equation}
where $\mathbf{H}_{\mathrm{DT}}$ is the delay-time domain channel matrix of size $MN \times MN$, $\mathbf{\Phi}_{\rm{DT}} = \mathrm{diag}(\boldsymbol{\psi}) $ is the phase noise matrix in the delay-time domain and 
\begin{equation}
    \boldsymbol{\psi} = [e^{j\theta[0]}, e^{j\theta[1]}, \dots , e^{j\theta[MN - 1 ]}]^{\rm{T}},
\end{equation}
is the vector containing the phase noise at each sampling point.

The received signal is then converted back to the delay-Doppler domain by a DFT along the time dimension, i.e.,
\begin{equation}
    \mathbf{y} = (\mathbf{F}_{N}\otimes\mathbf{I}_{M})\mathbf{r} + (\mathbf{F}_{N}\otimes\mathbf{I}_{M})\boldsymbol{\eta}.
\end{equation}
Using (2) and (5), (7) can be written as
\begin{equation}
    \mathbf{y} = \mathbf{G}_{\mathrm{DD}}\mathbf{x} + \boldsymbol{\eta}_{\mathrm{DD}}
    \label{eq8}
\end{equation}
where  $\mathbf{G}_{\mathrm{DD}} = \mathbf{\Phi}_{\rm{DD}}\mathbf{H}_{\mathrm{DD}}$
is the effective delay-Doppler domain channel matrix. The delay-Doppler domain phase noise matrix is given by $\mathbf{\Phi}_{\rm{DD}} = (\mathbf{F}_{N}\otimes\mathbf{I}_{M})\mathbf{\Phi}_{\rm{DT}}(\mathbf{F}_{N}^{\rm{H}}\otimes\mathbf{I}_{M})$ and the delay-Doppler domain representation of the LTV channel is given by $\mathbf{H}_{\mathrm{DD}} = (\mathbf{F}_{N}\otimes\mathbf{I}_{M})\mathbf{H}_{\mathrm{DT}}(\mathbf{F}_{N}^{\rm{H}}\otimes\mathbf{I}_{M}).$  The matrix $\mathbf{\Phi}_{\rm{DD}}$ is block circulent, i.e.,
\begin{equation}\mathbf{\Phi}_{\rm{DD}} = \left[
\begin{array}{cccc}
\mathbf{\Psi}_0 &\mathbf{\Psi}_{N-1} &\dots &\mathbf{\Psi}_{1}\\
\mathbf{\Psi}_1 &\mathbf{\Psi}_0 &{\dots} &\mathbf{\Psi}_{2} \\
\boldsymbol{\vdots} &\boldsymbol{\vdots} &{\ddots} &\boldsymbol{\vdots} \\
\mathbf{\Psi}_{N-1} &\mathbf{\Psi}_{N-2} &{\dots} &\mathbf{\Psi}_{0}
\end{array}
\right] \label{eq9}\end{equation}
where each $M\times M$ block matrix $\Psi_n$ has a diagonal structure and is given by
$\mathbf{\Psi}_{n} = \mathrm{diag}([\phi_n[0], \phi_n[1], \dots , \phi_{n}[M-1]]^{\rm{T}}).$
The delay-Doppler domain phase noise coefficients which make up the elements of $\mathbf{\Phi}_{\rm{DD}}$ are each given by
\setcounter{equation}{10}
\begin{equation}
    \phi_n[m] = \frac{1}{N}\sum_{i=0}^{N-1}e^{j\theta[m + nM]}e^{-j\pi n i/N}.
\end{equation}
The diagonal elements of $\mathbf{\Phi}_{\rm{DD}}$ are the delay-Doppler phase rotation components and the off-diagonal elements induce inter Doppler interference (IDI). In a large body of OFDM literature, phase noise estimation and compensation is only concerned with the common phase error rotation component and the ICI component is often ignored. An early OTFS paper on this topic,\cite{Bello_OTFS_pn}, adopted a similar approach. However, this approach assumes that the phase noise level is low and that the phase noise can be approximated as constant for each block of $M$ samples that makeup the OTFS transmission frame. In practical systems, this assumption may not hold. As of yet, there has been no detailed analysis of interference caused by phase noise in the delay-Doppler domain and the sensitivity of OTFS to this interference. Hence, this is the focus of the following section.

\section{Phase Noise Interference Analysis}
As mentioned above, the existing literature concludes that OTFS is more robust to phase noise than OFDM \cite{Surabhi_OTFS_PN_2019}. However, this conclusion generally assumes that phase noise is jointly estimated and compensated as a part of the channel. To the best of our knowledge, there is no prior work in the literature that analyses the sensitivity of OTFS to phase noise compared to OFDM in the absence of estimation and compensation. Thus, in this section, we analyze the sensitivity of OTFS to phase noise by providing a detailed SINR analysis of OTFS and OFDM over an AWGN channel with phase noise, which allows us to isolate the sole impact of phase noise on both systems. Hence, in this following, we consider an ideal channel, i.e., $\mathbf{H}_{\mathrm{DD}} = \mathbf{I}_{MN}$. Thus, we rewrite (\ref{eq8}) as 
\begin{equation}
    \mathbf{y} = \mathbf{\Phi}_{\rm{DD}} \mathbf{x} + \mathbf{w}.
\end{equation}
We note that the received signal vector $\mathbf{y}$ can be written as the concatenation of $N$ subvectors each containing $M$ elements
\begin{equation}
    \mathbf{y} = \left[ \mathbf{y}_0^{\mathrm{T}}, \mathbf{y}_1^{\mathrm{T}}, \dots, \mathbf{y}_{N-1}^{\mathrm{T}}  \right]^{\mathrm{T}},
\end{equation}
 where $\mathbf{y}_n = \left[ y_n[0], y_n[1], \dots y_n[M-1] \right]^{\mathrm{T}}$ represents the $n^{\rm th}$ delay block. From the structure of $\mathbf{\Phi}_{\rm{DD}}$ outlined in (\ref{eq10}), the input-output relationship for the received sample at element $m$ of $\mathbf{y}_n$ can be given as
 \begin{equation}
     y_n[m] = \phi_m[0]x_n[m] + \sum_{i=1}^{N-1} \phi_m[i]x_{(n-i)_N}[m] + w_n[m]. \label{eq14}
 \end{equation}
 The second term of (\ref{eq14}) represents the inter-Doppler interference (IDI) due to phase noise. Due to the statistical independence of the phase noise and the data symbols, the IDI power is obtained as
 \begin{equation}
     \sigma^2_{\mathrm{pn}} = \sum_{i=1}^{N-1}\mathbb{E}[|\phi_m[i]|^2]\mathbb{E}[|x_{(n-i)_N}[m]|^2].
 \end{equation}

Given the transmitted QAM symbols in $\mathbf{x}$, with unit average power, it follows that $\mathbb{E}[|x_{(n-i)_N}[m]|^2] = 1$. Therefore, to derive the interference power, we simply need to compute $\mathbb{E}[|\phi_m[i]|^2]$. This can be obtained from the diagonal elements of the delay-Doppler domain phase noise autocorrelation matrix, which we define as
\begin{equation}
    \boldsymbol{K}_{\phi_m} = \mathbb{E}[\boldsymbol{\phi}_m\boldsymbol{\phi}_m^{\mathrm{H}}],
\end{equation}
where 
$\boldsymbol{\phi}_m = [\phi_m[0], \phi_m[1], \dots , \phi_m[N-1]]^{\rm{T}}.$
The elements of $\boldsymbol{K}_{\phi_m}$ can be obtained as 
\begin{equation}
    \begin{split}
        &{K}_{\phi_m}[p,q] = \mathbb{E}\left[ \phi_{m}[p]\phi_{m}[q]\right]
        \\&= \mathbb{E}\left[\frac{1}{N^2}\sum_{k=0}^{N-1}\sum_{l=0}^{N-1}e^{j(\theta[m+kM] -\theta[m+lM])}e^{\frac{-j2\pi}{N}(pk-ql)}\right]
        \\& = \frac{1}{N^2}\sum_{k=0}^{N-1}\sum_{l=0}^{N-1}\mathbb{E}\left[e^{j\Delta\theta[{(k-l)M}]}\right]e^{\frac{-j2\pi}{N}(pk-ql)}.
    \end{split}\label{eq19}
\end{equation}
Since phase noise is a wide-sense stationary process, the expected value term is given by \cite{Fettweis_PN_OFDM}
\begin{equation}
    \mathbb{E}\left[e^{j\Delta\theta[{\delta}]}\right] = e^{-\frac{\sigma^2_{\theta}(\delta)}{2}}, \label{variogram}
\end{equation}
where $\sigma^2_{\theta}(\delta)$ is the time dependent variance or variogram of the phase noise process \cite{Chorti_PN_spectral_model}.
The enumeration of the variogram depends on the type of oscillator being used by the transmitter and receiver. Thus, for the sake of completeness, we consider three oscillator types, namely,  free-running oscillator, continuous-time PLL and discrete-time PLL in the following subsections.

\subsection{Free-Running Oscillator}
The simplest type of oscillator we can consider is a free-running oscillator. In this scenario, the phase noise is modeled as a Wiener process where the phase noise at time $n$ is given by
\begin{equation}
    \theta[n] = \theta[n-1] + \epsilon[n].
\end{equation}
This can also be represented via a linear transfer function in the z-domain as
\begin{equation}
    H_{\rm{W}}(z) = \frac{z}{z-1}.
    \label{H_w(z)}
\end{equation}
The phase error at time $n$ is modeled as
$$ \epsilon[n] \sim \mathcal{N}(0, \nu^2_{\mathrm{pn}}),  $$
and $\nu^2_{\mathrm{pn}}$ is the phase noise variance from sample to sample. For a free-running oscillator the phase noise variance is given by
$$\nu^2_{\mathrm{pn}} = 4\pi\beta_{\mathrm{pn}} T_{\rm{s}}, $$
where $\beta_{\mathrm{pn}}$ is the one-sided 3~dB line width of the oscillator's Lorentzian power spectral density.
In this scenario, the delay-dependent variance is given by
\begin{equation}
    \sigma^2_{\delta}(\delta) = 4\pi\beta_{\mathrm{pn}} T_{\rm{s}} \delta.\label{FRO_var}
\end{equation}
Hence, for a free-running oscillator, the expected value term in (\ref{eq19}) is obtained as
\begin{equation}
    \mathbb{E}\left[e^{j\Delta\theta[{(k-l)M}]}\right] = e^{-2\pi\beta_{\rm{pn}}T_{\rm{s}}|k-l|M}. \label{E_fro}
\end{equation}

\subsection{Continuous-time Phase Locked Loop}

\begin{figure}[t]
    \centering
\includegraphics[width=\linewidth]{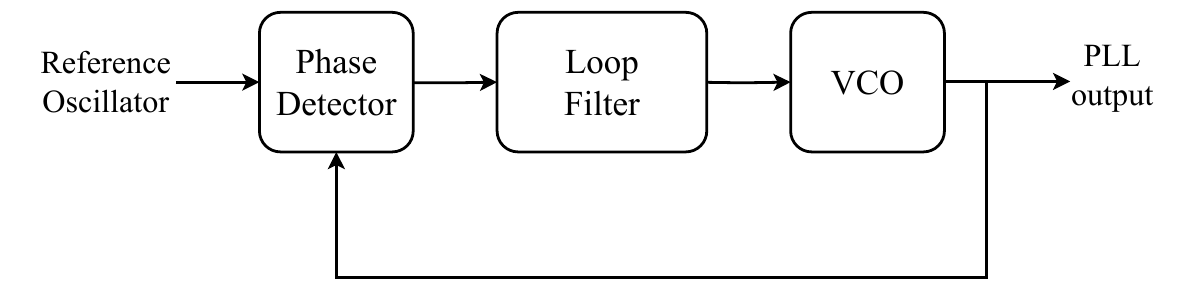}
    \caption{Block diagram illustration of generic PLL set up including a reference oscillator, phase detector, loop filter and VCO}
    \label{fig:PLL_fig}
\end{figure}

\begin{figure}[t]
    \centering
\includegraphics[width=\columnwidth]{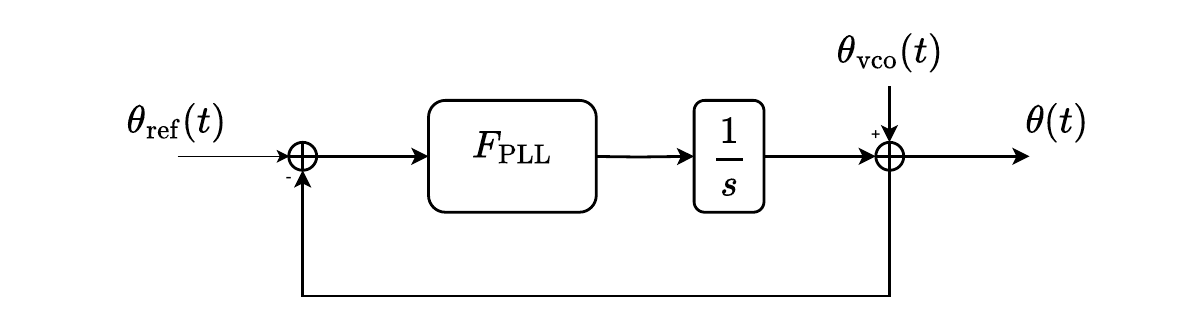}
    \caption{ Block diagram illustration of a First order continuous-time PLL diagram }
    \label{fig:contPLL_fig}
\end{figure}

In practical systems a PLL control system is used to generate a more stable oscillator output signal than a simple  free-running oscillator. A PLL consists of a negative feedback loop surrounding a low-quality voltage controlled oscillator (VCO) who's frequency is controlled by a high-quality reference signal, a phase detector and a low-pass filter, as can be seen in Fig. \ref{fig:PLL_fig}. In this paper, we consider a first order continuous-time PLL, see the block diagram in Fig \ref{fig:contPLL_fig}. The system consists of a reference signal, a VCO and a filter with filter coefficient $F_{\rm{PLL}}$. For this work, we assume a noiseless reference oscillator and noisy VCO oscillator. The transfer function for VCO phase noise to the oscillator output is given by 
\begin{equation}
    H(s) = \frac{s}{s+F_{\mathrm{PLL}}},
    \label{H(s)}
\end{equation}
which is a high-pass filter. This output phase noise is modeled as an Ornstein-Uhlenbeck process as opposed to a Wiener process \cite{Fettweis_PN_OFDM}. Assuming Wiener noise  with zero mean and variance $\nu^2_{\mathrm{pn}}$ at the VCO then the variogram for the continuous-time PLL is given by \cite{Demir_PN_2006}
\begin{equation}
    \sigma^2_{\theta}(\delta) = \frac{2\pi\beta_{\rm{pn}}}{F_{\mathrm{PLL}}}(1-e^{(-\delta{F_{\mathrm{PLL}}T_{\rm{s}})}}).
\end{equation}
Hence, for a continuous-time PLL, the expected value term in (\ref{eq19}) is obtained as
\begin{equation}
    \mathbb{E}\left[e^{j\Delta\theta[{(k-l)M}]}\right] = e^{-\frac{\pi\beta_{\rm{pn}}}{F_{\mathrm{PLL}}}(1-e^{(-|k-l|M{F_{\mathrm{PLL}}T_{\rm{s}})}})}. \label{E_cpll}
\end{equation}

\subsection{Discrete-time Phase Locked Loop}

\begin{figure}[t]
    \centering
\includegraphics[width=\columnwidth]{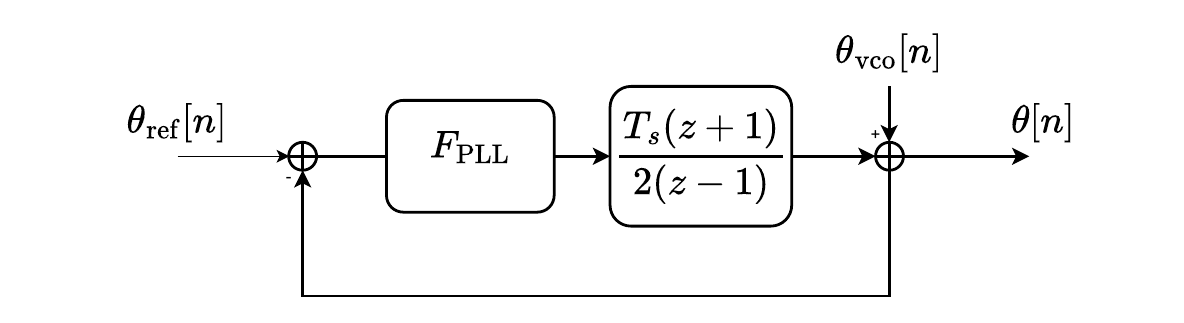}
    \caption{Block diagram illustration of a First order discrete-time PLL diagram}
    \label{fig:discPLL_fig}
\end{figure}

In this section, we consider the discrete-time PLL equivalent of the continuous-time PLL, which is shown in Fig. \ref{fig:discPLL_fig}. As one of the contributions of this work, we derive the variogram for the discrete-time PLL which, to the best of our knowledge, has not been previously reported in the literature. The discrete-time version of the PLL outlined above is given by using the bilinear transform to transform from the $s$-domain to the $z$-domain. By substituting 
$s = \frac{2}{T_{\rm{s}}}\frac{z-1}{z+1}$ 
into (\ref{H(s)}), we obtain the discrete-time transfer function of the system,
\begin{equation}
    H_{\rm{PLL}}(z) = \frac{b(z-1)}{z-a},
    \label{H(z)}
\end{equation}
where
\begin{equation}
    b = \frac{2}{2+T_{\rm{s}} F_{\mathrm{PLL}}}
    \label{b_pll}
\end{equation}
and
\begin{equation}
    a = \frac{2-T_{\rm{s}}F_{\mathrm{PLL}}}{2+T_{\rm{s}} F_{\mathrm{PLL}}}.
    \label{a_pll}
\end{equation}
The variogram of the discrete-time PLL system can be calculated as 
\begin{equation}
    \sigma^2_{\theta}[n] = K_{\theta}[0] - K_{\theta}[n],
\end{equation}
where $K_{\theta}[n]$ is the autocorrelation of the time domain PLL phase noise process at sample point $n$. Since the variogram is a positive and even function, the autocorrelation is given via the Wiener–Khinchin theorem as \cite{Chorti_PN_spectral_model} 
\begin{equation}
    K_{\theta}[n] = \frac{1}{\pi}\int_{0}^{\pi}|H_{\rm{PLL}}(\omega)|^2|H_{\rm{W}}(\omega)|^2S_{\epsilon}(\omega)\cos{(2\omega n)}\,d\omega,
    \label{R_theta}
\end{equation}
where $S_{\nu}(\omega)$ is the power-spectral density of the white noise source at the VCO output. Using (\ref{H(z)}) and (\ref{H_w(z)}), we can expand (\ref{R_theta}) as

\begin{equation}
\begin{split}
    K_{\theta}[n] &= \frac{1}{\pi}\int_{0}^{\pi}\left|\frac{b(e^{j\omega}-1)}{e^{j\omega}-a}\right|^2\left|\frac{e^{j\omega}}{e^{j\omega}-1}\right|^2S_{\epsilon}(\omega)\cos{(2\omega n)}\,d\omega,
    \\&= \frac{1}{\pi}\int_{0}^{\pi}\frac{b^2(2-2\cos{(\omega)})}{1+a-2a\cos{(\omega)}}\frac{\nu^2_{\mathrm{pn}}\cos{(2\omega n)}}{(2-2\cos{(\omega)})}\,d\omega,
    \\&= \frac{b^2 \nu^2_{\mathrm{pn}}}{\pi}\int_{0}^{\pi}\frac{\cos{(2\omega n)}}{1+a-2a\cos{(\omega)}} \,d\omega
    \\&= \frac{b^2 \nu^2_{\mathrm{pn}}}{1-a^2}a^{|2n|}.
\end{split}
\end{equation}
We note that this is the autocorrelation function of a discrete-time AR(1) process which is the discrete-time equivalent of a  continuous-time domain Ornstein-Uhlenbeck process \cite{OU_AR1}. Accordingly, the variogram of the discrete-time PLL system is obtained as
\begin{equation}
    \sigma^2_{\theta}(\delta) = \frac{b^2 \nu^2_{\mathrm{pn}}}{1-a^2}(1-a^{|2\delta|}) \label{vgram_dpll}.
\end{equation}
By using (\ref{b_pll}) and (\ref{a_pll}), we can rewrite (\ref{vgram_dpll}) as
\begin{equation}
    \sigma^2_{\theta}(\delta) = \frac{2\pi\beta_{\rm{pn}}}{F_{\mathrm{PLL}}}\left(1-\left(\frac{2-T_{\rm{s}}F_{\mathrm{PLL}}}{2+T_{\rm{s}} F_{\mathrm{PLL}}}\right)^{|2\delta|}\right),
\end{equation}
Hence, for a first order discrete-time PLL, the expected value term in (\ref{eq19}) is obtained as
\begin{equation}
    \mathbb{E}\left[e^{j\Delta\theta[{(k-l)M}]}\right] = e^{\frac{\pi\beta_{\rm{pn}}}{F_{\mathrm{PLL}}}\left(1-\left(\frac{2-T_{\rm{s}}F_{\mathrm{PLL}}}{2+T_{\rm{s}} F_{\mathrm{PLL}}}\right)^{2M|k-l|}\right)}. \label{E_dpll}
\end{equation}

\subsection{Signal to Interference Plus Noise Ratio}
In this subsection, we utilize the analysis from subsections A - C to derive an expression for the SINR at the receiver.
The SINR at the receiver is given by \cite{Fettweis_PN_OFDM}
\begin{equation}
    \mathrm{SINR} = \frac{{K}_{\phi}[0,0]}{\sigma^2_{\mathrm{pn}} + \sigma^2_{\eta}}.
\end{equation}
The IDI power due to phase noise can be calculated using the diagonal elements of $\boldsymbol{K}_{\phi_m}$ as
\begin{equation}
    \sigma^2_{\mathrm{pn}} =  \sum_{p=1}^{N-1}{K}_{\phi}[p,p],
\end{equation}
where the exact expression for this interference depends on the type of oscillator utilized in the system. 

For a free-running oscillator, using (\ref{E_fro}), we can obtain the diagonal elements of $\boldsymbol{K}_{\phi_m}$ as
\begin{equation}
   {K}^{\mathtt{FRO}}_{\phi_m}[p,p] = \frac{1}{N^2}\sum_{k=0}^{N-1}\sum_{l=0}^{N-1}e^{-2\pi\beta_{\rm{pn}}T_{\rm{s}}|k-l|M}e^{\frac{-j2\pi}{N}(p(k-l))}. \label{Kpp_fro}
\end{equation}
Next, we will derive the closed form expression of (\ref{Kpp_fro}). First, by defining $n=k-l$, we can simplify the expression in (\ref{Kpp_fro}) as
\begin{equation}
   {K}^{\mathtt{FRO}}_{\phi_m}[p,p] = \frac{1}{N^2}\sum_{n=-N+1}^{N-1}(N-|n|)e^{-2\pi\beta_{\rm{pn}}T_{\rm{s}}|n|M}e^{\frac{-j2\pi p}{N}n},
\end{equation}
which can be rewritten as
\begin{equation}
\begin{split}
   {K}^{\mathtt{FRO}}_{\phi_m}[p,p] = &\frac{1}{N}\sum_{n=-N+1}^{N-1}e^{-2\pi\beta_{\rm{pn}}T_{\rm{s}}|n|M}e^{\frac{-j2\pi p}{N}n}
   \\&-\frac{1}{N^2}\sum_{n=-N+1}^{N-1}|n|e^{-2\pi\beta_{\rm{pn}}T_{\rm{s}}|n|M}e^{\frac{-j2\pi p}{N}n}. \label{Kpp_fro2}
   \end{split}
\end{equation}
Next, we can remove the absolute value functions in the summations in (\ref{Kpp_fro2}) by utilizing $\sum_{n=-N-1}^{N-1}e^{a|n|}e^{bn} = \sum_{n=0}^{N-1}(e^{(a+b)n}+e^{(a-b)n}) - 1$. Hence, we can rewrite (\ref{Kpp_fro2}) as 
\begin{equation}
\begin{split}
   &{K}^{\mathtt{FRO}}_{\phi_m}[p,p] =
   \\&\frac{1}{N}\left[\sum_{n=0}^{N-1}(e^{-(2\pi\beta_{\rm{pn}}T_{\rm{s}}M + \frac{-j2\pi p}{N})n} + e^{-(2\pi\beta_{\rm{pn}}T_{\rm{s}}M - \frac{-j2\pi p}{N})n}) -1\right]\\
   &-{\frac{1}{N^2}}\!\!\!\sum_{n=1}^{N-1}\left(ne^{-(2\pi\beta_{\rm{pn}}T_{\rm{s}}M + \frac{-j2\pi p}{N})n} + ne^{-(2\pi\beta_{\rm{pn}}T_{\rm{s}}M - \frac{-j2\pi p}{N})n}\right)
   \end{split}
\end{equation}

Finally, by defining $\alpha = e^{-2\pi\beta_{\rm{pn}}T_{\rm{s}}M}$ and utilizing $\sum_{n=0}^{N-1}z^{n} = \frac{1-z^{N}}{1-z^{}}$ and $\sum_{n=1}^{N-1}nz^{n} = \frac{z-Nz^{N} + (N-1)z^{(N+1)}}{(1-z)^2},$ the closed form expression for the diagonal elements of $\boldsymbol{K}^{\mathtt{FRO}}_{\phi_m}$ is obtained as
\begin{equation}
    \begin{split}
    &{K}^{\mathtt{FRO}}_{\phi_m}[p,p] = \frac{1}{N} \left[ \frac{2\alpha^2\cos{(\frac{2 \pi p}{N})-2\alpha +1}}{1-2\alpha\cos{(\frac{2 \pi p}{N})} + \alpha^2} \right]- \\&
    \frac{1}{N^2} \left[\frac{2(1-2\alpha - \alpha^2)\cos{(\frac{2 \pi p}{N})} - 4\alpha^2\cos^2{(\frac{2 \pi p}{N})} -2\alpha}{(1-2\alpha\cos{(\frac{2 \pi p}{N})} + \alpha^2)^2}  \right].
     \label{K_pp_fro_FINAL}
    \end{split}
\end{equation}

For the case where continuous-time PLL is used, we can obtain the diagonal elements of $\boldsymbol{K}_{\phi_m}$, using (\ref{E_cpll}), as
\begin{equation}
\begin{split}
   &{K}^{\mathtt{CPLL}}_{\phi_m}[p,p] = 
   \\&\frac{1}{N^2}\sum_{k=0}^{N-1}\sum_{l=0}^{N-1}e^{-\frac{\pi\beta_{\rm{pn}}}{F_{\mathrm{PLL}}}(1-e^{(-(k-l|M{F_{\mathrm{PLL}}T_{\rm{s}})}})}e^{\frac{-j2\pi}{N}(p(k-l))}
   \end{split}
\end{equation}
This expression can be further simplified by defining $n=k-l$ and rewriting as
\begin{equation}
\begin{split}
    &{K}^{\mathtt{CPLL}}_{\phi_m}[p,p]= 
    \\& \frac{1}{N^2}\!\!\!\!\sum_{n=-N+1}^{N-1}\!\!\!\!\!(N-|n|)e^{-\frac{\pi\beta_{\rm{pn}}}{F_{\mathrm{PLL}}}(1-e^{(-|n|M{F_{\mathrm{PLL}}T_{\rm{s}})}})}e^{\frac{-j2\pi p}{N}n}\label{Kpp_cpll1} 
\end{split}
\end{equation}
Since the expression in (\ref{Kpp_cpll1}) contains a double exponential term, it is not possible to obtain a closed form expression to it, as was done in the case of a free-running oscillator.

Finally, for a discrete-time PLL, we use (\ref{E_dpll}) to obtain the diagonal elements of $\boldsymbol{K}_{\phi_m}$ as
\begin{equation}
\begin{split}
    &{K}^{\mathtt{DPLL}}_{\phi_m}[p,p] =\\& \frac{1}{N^2}\sum_{k=0}^{N-1}\sum_{l=0}^{N-1}e^{\frac{\pi\beta_{\rm{pn}}}{F_{\mathrm{PLL}}}\left(1-\left(\frac{2-T_{\rm{s}}F_{\mathrm{PLL}}}{2+T_{\rm{s}} F_{\mathrm{PLL}}}\right)^{2M|k-l|}\right)}e^{\frac{-j2\pi}{N}(p(k-l))}, \label{Kpp_dpll1}
\end{split}
\end{equation}
which can be simplified by defining $n=k-l$ and rewriting as
\begin{equation}
   {K}^{\mathtt{DPLL}}_{\phi_m}[p,p] = \frac{1}{N^2}\!\!\!\!\sum_{n=-N+1}^{N-1}\!\!\!\!(N-|n|)e^{\frac{\pi\beta_{\rm{pn}}}{F_{\mathrm{PLL}}}(1-a^{2|n|M})}e^{\frac{-j2\pi p}{N}n}.\label{Kpp_dpll1}
\end{equation}
As with the continuous-time PLL case, the presence of a double exponential term in (\ref{Kpp_dpll1}) means that it is not possible to obtain a closed form expression.

\begin{figure}[t]
    \centering
    \includegraphics[width=0.9\columnwidth]{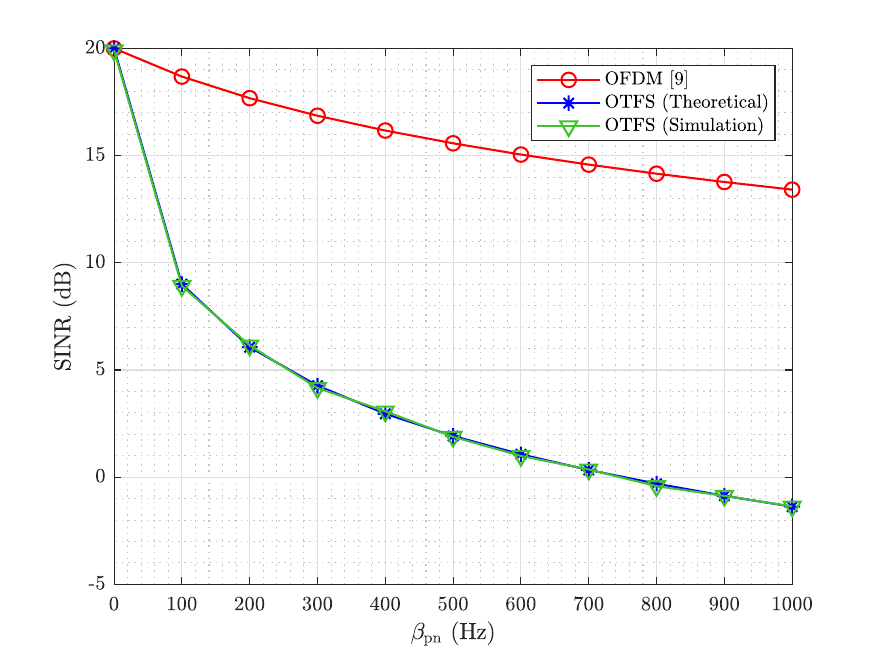}
    \caption{SINR of OFDM and OTFS in the presence of phase noise compared to the derived theoretical expression as phase noise bandwidth increases}
    \label{fig:results1.4}
\end{figure}

Next, we present simulation results to analyze the sensitivity of OTFS to phase noise in comparison to OFDM when phase noise is not estimated or compensated. A carrier frequency of $f_\mathrm{c} = 5.9$ GHz, a transmission bandwidth of 7.68~MHz and a delay-Doppler grid size of $M=128$ and $N=32$ are considered for this simulation. In Fig. \ref{fig:results1.4}, we compare our derived expression for a  free-running oscillator system to measured values obtained via Monte Carlo simulation for increasing values of $\beta_{\rm{pn}}$. Additionally, we compare the OTFS SINR to the SINR of equivalent OFDM system using the expressions derived in \cite{Fettweis_PN_OFDM}. It can be seen in Fig.~5 that when phase noise bandwidth increases from 0~Hz to 100~Hz, the SINR degradation for OFDM is approximately 1~dB whereas OTFS experiences a degradation exceeding 10~dB. This result highlights that OTFS is significantly more sensitive to phase noise than OFDM when no estimation and compensation is applied. To illustrate why this is the case, we will now present the interference expression for an equivalent OFDM system with $M$ subcarriers and $N$ OFDM symbols. From \cite{Fettweis_PN_OFDM}, the frequency domain interference power due to phase noise is given by 
\begin{equation}
    \sigma^2_{\mathrm{fpn}} =  \sum_{p=1}^{M-1}{K}_{\varphi}[p,p],
\end{equation}
where $\boldsymbol{K}_{\varphi}$ is the frequency domain autocorrelation matrix given by \begin{equation}
    \boldsymbol{K}_{\varphi} = \mathbb{E}[\boldsymbol{\varphi}\boldsymbol{\varphi}^{\mathrm{H}}].
\end{equation}
 The $M\times 1$ vector $\boldsymbol{\varphi}$ contains the frequency domain phase noise coefficients which are given by
 \begin{equation}
    \varphi[m] = \frac{1}{M}\sum_{i=0}^{M-1}e^{j\theta[i]}e^{-j\pi m i/M}.
\end{equation}
We can now obtain the elements of $\boldsymbol{K}_{\varphi}$ as 
\begin{equation}
    \begin{split}
        &{K}_{\varphi}[p,q] = \mathbb{E}\left[ \varphi[p]\varphi[q]\right]
        \\& = \frac{1}{M^2}\sum_{k=0}^{M-1}\sum_{l=0}^{M-1}\mathbb{E}\left[e^{j\Delta\theta[{(k-l)}]}\right]e^{\frac{-j2\pi}{M}(pk-ql)}.
    \end{split}
\end{equation}
For the case of a free running oscillator, by using (\ref{variogram}) and (\ref{FRO_var}), we can obtain the diagonal elements of $\boldsymbol{K}_{\varphi}$ as
\begin{equation}
   {K}^{\mathtt{FRO}}_{\varphi}[p,p] = \frac{1}{M^2}\sum_{k=0}^{M-1}\sum_{l=0}^{M-1}e^{-2\pi\beta_{\rm{pn}}T_{\rm{s}}|k-l|}e^{\frac{-j2\pi}{M}(p(k-l))}. \label{autocorr_freq}
\end{equation} 
It can be seen from (\ref{autocorr_freq}) and (\ref{Kpp_fro}) that the main difference between the OTFS and OFDM systems is the the factor of $M$ present in the exponent in (\ref{Kpp_fro}). This factor of $M$ appears because the row-wise DFT operation of OTFS takes signal samples which are spaced apart by $MT_{\rm{s}}$ seconds and brings them to the delay-Doppler domain. In contrast, OFDM performs the DFT on contiguous samples which are spaced apart by $T_{\rm{s}}$ seconds. This increased spacing leads to higher sample-to-sample variance in the delay-Doppler domain and leads to increased interference. This can also be thought of as being analogous to OFDM in the presence of phase noise process with a sample-to-sample variance, i.e., $\nu^2_{\mathrm{pn}}$, which is $M$ times larger.

 This analysis demonstrates that OTFS is more sensitive to phase noise than OFDM when phase noise is not estimated and compensated. While OFDM receivers do not need to deal with phase noise until it is high and can use phase tracking reference pilots for CPE estimation and compensation, our analysis shows that OTFS experiences much greater IDI at lower levels of phase noise. This provides intuition as to why a CPE estimation approach, which ignores IDI, such as the method proposed in [13] is not appropriate for OTFS systems. The analysis and derivations presented in this section pave the way towards our proposed joint phase noise and channel estimation technique which will be presented in the following section. It should be noted that with perfect knowledge of the instantaneous phase noise process and an optimal receiver, the channel capacity of OTFS and OFDM under phase noise will be identical. However, when considering practical, sub-optimal receivers, such as MMSE, and imperfect estimation of the phase noise, this increased sensitivity to phase noise could greatly effect the performance of OTFS. The analysis and derivations presented in this section pave the way towards our proposed joint phase noise and channel estimation technique which will be presented in the following section.

\section{Proposed Joint Phase Noise and Channel Estimation Technique}
In this section, we outline our proposed joint phase noise and channel estimation technique for OTFS systems. Additionally, we provide a detailed computational complexity analysis of the proposed technique in terms of number of complex multiplications and compare it to existing methods in the literature. The proposed technique uses the statistical nature of the phase noise and the Doppler spread to estimate the effective channel. The proposed technique has 2 stages. In the first stage, an isolated impulse pilot in delay-Doppler domain, which translates into a train of equally spaced impulses in the delay-time domain, is used to estimate the channel at the pilot indices. The second stage uses Wiener filtering, based on the statistical nature of the phase noise and Doppler spread processes, to estimate the missing channel samples between the delay-time impulse pilot estimates.

\subsection*{Stage 1: Partial channel estimation}
Since the procedure outlined later in stage 2 takes place in the delay-time domain, in stage 1, we estimate the channel in the delay-time domain. We place a strong impulse pilot at the indices $(m_p, n_p)$ on the delay-Doppler grid. The pilot is then surrounded by zero guards up to $L-1$ delay bins above and below to absorb interference from the data to the pilot due to the delay spread. When converted to the delay-time domain, the isolated delay-Doppler domain pilot will have an impulse train representation in the delay-time domain with indices that are equally spaced apart with a spacing of $M$ samples.

\begin{figure}[t]
    \centering
    \begin{subfigure}{0.49\columnwidth}
    \includegraphics[width=\textwidth]{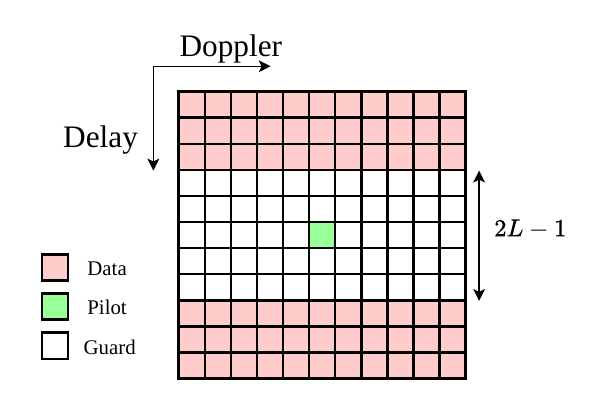}
    \caption{Pilot in delay-Doppler grid.}
    \label{fig:impulse1}
    \end{subfigure}
    \hfill
    \begin{subfigure}{0.49\columnwidth}
    \includegraphics[width=\textwidth]{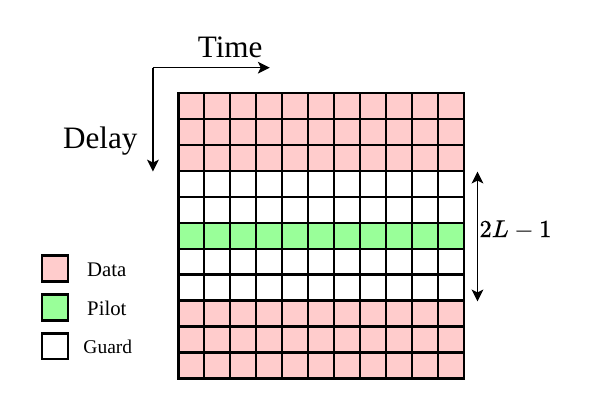}
    \caption{Pilot in delay-time grid.}
    \label{fig:impulse2}
    \end{subfigure}
    \caption{Illustration of the delay-time domain and delay-Doppler domain pilot pattern used for obtaining the initial estimate of the channel and phase noise.}
\end{figure}

We define the set of delay-time pilot indices as $\mathcal{P} \!\!=\!\! \{{m_p}, {m_p+M}, \dots, {m_p \!+\!M(N-1)}\}$. This pilot structure, in the delay-time domain, is illustrated in Fig. \ref{fig:impulse2}. We then use a threshold based method to estimate each of the channel delay taps at the pilot indices \cite{Raviteja_OTFS_2019}. The delay-time channel impulse response is estimated at the pilot region by capturing the $L \times N$  matrix at the delay bins at indices $\{m_p, m_p \!+\!1, \dots ,m_p \!+\!L-1\}$. This estimated channel impulse response matrix is given by
\begin{equation}
    \widehat{\mathbf{G}}_{\mathrm{DT}}^{\mathcal{P}} = [\widehat{\mathbf{g}}_{\rm{0}}, \widehat{\mathbf{g}}_{\rm{1}}, \dots, \widehat{\mathbf{g}}_{{L-1}}]^{\rm{T}},
\end{equation}
 where $\widehat{\mathbf{g}}_{{l}}$ is a $N\times1$ vector containing the channel estimates of the $l$-th channel delay tap at each of the pilot indices in $\mathcal{P}$. This stage only obtains the channel estimate at pilot location on the delay-time grid. However, as the actual channel is varying from sample-to-sample due to the phase noise and Doppler spread, there are missing channel samples from in between the pilots. Therefore, at the next stage, we propose a statistical approach to obtain the sample-to-sample channel estimate, without the need for an increase in pilot overhead.

\subsection*{Stage 2: Full Channel Estimation} 
 Possible approaches to finding the missing channel estimates in the OTFS literature include BEM and Spline interpolation \cite{Sanoop_PCP, Thaj_OTSM_2021}. As stated earlier, these methods are well suited for low-pass processes such as Doppler spread channels as the variations of the channel are smooth sample-to-sample. However, since phase noise is a wideband process and the variations are not smooth sample-to-sample, BEM and Spline do not provide satisfactory performance in phase noise estimation.

Hence, in the following, we propose a Wiener filter channel estimation technique based on the statistical nature of both the Doppler spread channel and the phase noise to improve upon the partial channel estimate from stage 1. The aim of our proposed channel estimation technique is to design a filter matrix $\mathbf{W}$ which minimizes the mean-squared error between the estimated channel and actual effective channel for each of the $L$ channel taps. 
We can write this problem as 
\begin{equation}     
            \min_{\mathbf{W}}    \qquad \mathbb{E}[|\mathbf{W}\widehat{\mathbf{g}}_{\rm{l}} - \mathbf{g}_l|^2],\label{W_min}
\end{equation}
where $\mathbf{g} = [ \psi[0]h[0,l], ... ,\psi[MN-1]h[MN-1,l]]^{\rm{T}}$ is the vector containing the effective channel samples of channel tap $l$. Since each delay tap experiences the same phase noise and Doppler spread statistics, (\ref{W_min}) leads to the same result for all values of $l$ and the only difference between the taps is the tap gain which scales the whole equation. Thus, we only need to solve (\ref{W_min}) for one delay tap and can apply the solution to each column to each column of $\widehat{\mathbf{G}}_{\mathrm{DT}}^{\mathcal{P}}$ as they are already scaled by the tap gains. Hence, without loss of generality, we we omit the subscript $l$ in the following.  
The solution to (\ref{W_min}) can be obtained via solving the Wiener-Hopf equations \cite{Adaptive_filters_book} and is given by 
\begin{equation}
    \mathbf{W} = \boldsymbol{K}_{g,\widehat{g}}(\boldsymbol{K}_{\widehat{g},\widehat{g}})^{\dag}
\end{equation}
where $\boldsymbol{K}_{g,\widehat{g}} = \mathbb{E}\left[\mathbf{g}\widehat{\mathbf{g}}^{\rm{H}} \right]$ is the size $MN\times N$ cross-correlation matrix of the effective channel and the estimated channel at the pilot indices. The matrix $\boldsymbol{K}_{\widehat{g},\widehat{g}} = \mathbb{E}\left[\widehat{\mathbf{g}}\widehat{\mathbf{g}}^{\mathrm{H}} \right]$ is the size $N \times N$ autocorrelation matrix of the estimated channel at the pilot indices and is given by
\begin{equation}
\begin{split}
    \boldsymbol{K}_{\widehat{g},\widehat{g}} &= \mathbb{E}\left[\widehat{\mathbf{g}}\widehat{\mathbf{g}}^{\rm{H}} \right]\\
    &= \left(\boldsymbol{K}_{g,g}^\mathcal{P} + \frac{\sigma_\eta^2}{\sigma_{\mathrm{p}}^2}\mathbf{I}_N\right).
\end{split} \label{K_gg}   
\end{equation}
In (\ref{K_gg}), $\boldsymbol{K}_{g,g}^\mathcal{P}$ is the effective channel autocorrelation matrix at the pilot indices and $\sigma_{\mathrm{p}}^2$ represents the pilot power. The elements of $\boldsymbol{K}_{g,\widehat{g}}$ and $\boldsymbol{K}_{g,g}^\mathcal{P}$ are obtained by 
$${K}_{g,\widehat{g}}[m,n] = {K}_{g}[m,p_n],$$
 and 
 $${K}_{g,g}^\mathcal{P}[m,n] = {K}_{g}[p_m,p_n],$$
 respectively, where $\boldsymbol{K}_\mathrm{g}$ is the $MN \times MN$ autocorrelation matrix of the full effective channel.
 
Since the Doppler spread and the oscillator phase noise are statistically independent, $\boldsymbol{K}_\mathrm{g}$ can be obtained via the product of their autocorrelation of matrices
\begin{equation}
    \boldsymbol{K}_\mathrm{g} =  \boldsymbol{K}_{\psi}\boldsymbol{K}_\mathrm{D},
\end{equation}
where $\boldsymbol{K}_\mathrm{D}$ is the autocorrelation matrix of the Doppler spread channel and $\boldsymbol{K}_{\psi}= \mathbb{E}\left[\boldsymbol{\psi}\boldsymbol{\psi}^{\rm{H}} \right]$ is the autocorrelation matrix of the phase noise process. Assuming the Doppler effect follows the well-known Jakes model \cite{jakes_model}, the elements of $\boldsymbol{K}_\mathrm{D}$ are given by 
\begin{equation}
    {K}_\mathrm{D}[m,n] =  J_0(2\pi f_\mathrm{D}T_{\rm{s}}|m-n|)
\end{equation}
where $J_0(\cdot)$ is the zeroth order Bessel function of the first kind and $f_\mathrm{D}$ is the maximum Doppler shift in the channel. 

For the phase noise autocorrelation matrix, the elements of $\boldsymbol{K}_{\psi}$ are given by
\begin{equation}
\begin{split}
    {K}_{\psi}[m,n] &= \mathbb{E}\left[ \psi[m]\psi^{*}[n]\right] \\
    &= \mathbb{E}\left[e^{j\Delta\theta[{(m-n)}]}\right],\\
    &=  e^{-\frac{\sigma^2_{\theta}(|m-n|)}{2}},\\
\end{split} 
\label{K_psi}
\end{equation}
The analysis from Section~III is then directly used to compute the elements of $\boldsymbol{K}_{\psi}$ using the variogram function of the phase noise process for 3 different oscillator types. Since we are concerned with the delay-time domain for the purposes of estimation,  we obtain (\ref{K_psi}) for each oscillator type as
\begin{equation}
{K}_{\psi}[m,n] =
\left\{ 
    \begin{array}{lll}
        e^{-2\pi\beta_{\rm{pn}}T_{\rm{s}}|m-n|},&\qquad \mathrm{FRO}  &\\
        e^{-\frac{\pi\beta_{\rm{pn}}}{F_{\mathrm{PLL}}}(1-e^{(-|m-n|{F_{\mathrm{PLL}}T_{\rm{s}})}})}, &\qquad \mathrm{CPLL} &\\
        e^{\frac{\pi\beta_{\rm{pn}}}{F_{\mathrm{PLL}}}(1-a^{2|m-n|})}, &\qquad \mathrm{DPLL}.
    \end{array} 
\right. \label{K_psi2}
\end{equation}
It should be noted that the expressions outlined in (\ref{K_psi2}) apply only to the case where white noise sources are present in the oscillator. In the presence of flicker noise sources, closed-form expressions of the auto-correlation function are difficult, if not impossible, to obtain. However, the proposed method can still be applied for phase noise estimation for oscillators exhibiting flicker noise. The auto-correlation function of the phase noise can be obtained from the oscillator power spectral density function which itself can be obtained either from manufacturer specifications \cite{Demir_PN_2006} or via phase noise spectral modeling \cite{collmann2025practicalanalysisunderstandingphase}.

The full delay-time domain channel estimate is obtained by multiplying the Wiener filtering matrix by the initial snapshot estimate from stage 1,
\begin{equation}
    \bG_{\rm DT}^{\mathtt{W}} = \mathbf{W}\widehat{\mathbf{G}}_{\mathrm{TD}}^{\rm{P}}.
\end{equation}
This estimate can be used to perform linear equalization techniques in the delay-time domain techniques after which the equalized data data can be converted to the delay-Doppler domain for detection. Alternatively, the delay-Doppler domain channel estimate can be obtained via
\begin{equation}
    \bG_{\rm DD}^{\mathtt{W}} = (\mathbf{F}_{N}\otimes\mathbf{I}_{M})\bG_{\rm DT}^{\mathtt{W}}(\mathbf{F}_{N}^{\rm{H}}\otimes\mathbf{I}_{M}),
\end{equation}
and equalization and detection can be performed in that domain.

\subsection{Computational complexity}
In this section, we compare the computational complexity of the proposed method to existing methods in the literature. Firstly, the computation of the Wiener filter matrix $\mathbf{W}$ is performed offline based on the channel and phase noise statistics. For the online portion of the proposed method, stage 1 utilizes the threshold based method from \cite{Raviteja_OTFS_2019} which has a complexity of $\mathcal{O}(LN)$. In Stage 2 of the proposed technique, computing $\bG_{\rm DT}^{\mathtt{W}}$ has a complexity of $\mathcal{O}(MN^2L)$. In comparison, BEM based interpolation has a complexity of $\mathcal{O}(LNQ_{\rm{BEM}}(M+1))$, where $Q_{\rm{BEM}}=\lceil 2k_\mathrm{over}M_{\rm T}N(f_{\rm D}+\beta_{\rm{pn}})T_{\rm s}\rceil+1$ is the number of BEM basis functions used. It is clear that the complexity of the BEM based estimation depends upon the severity of the phase noise. When phase noise is low and $Q_{\rm{BEM}}<N$, BEM based estimation can have lower complexity than the proposed method. However, when the phase noise bandwidth is larger and $Q_{\rm{BEM}}>N$, the proposed method has lower complexity as it does not depend on the phase noise bandwidth. Meanwhile, Spline based estimation has a computational complexity of $\mathcal{O}(2(N-1)ML)$ \cite{Thaj_OTSM_2021}, which is lower than the complexity of both BEM and the proposed method. However, as will be shown in the numerical results presented in the following section, Spline is ineffective in the presence of phase noise and provides negligible performance improvement over using Stage 1 of the proposed technique.

\section{Numerical results}

\begin{table}[t]
    \caption{Simulation Parameters}
    \centering
    \begin{tabular}{|c|c|}
        \hline
         Delay bins ($M$) & 128\\
         \hline
         Doppler bins ($N$) & 32 \\
         \hline
         Carrier frequency ($f_c$) & 5.9 GHz \\
         \hline
         Subcarrier spacing  & 60 kHz \\
         \hline
         Modulation scheme & 4-QAM, 16-QAM  \\
         \hline
         Channel model & TDL-C \cite{3gpp_TS38901} \\
         \hline
         Delay spread & 100 ns  \\
         \hline
         Velocity & $0$km/h, $500$ km/h  \\
         \hline
         $\beta_{\rm{PN}}$ & $2\times10^3$ Hz  \\
         \hline
         $F_{\rm{PLL}}$ & $1\times10^6$  \\
         \hline
    \end{tabular}
    \label{tab:table 1}
\end{table}

This section presents numerical results to showcase the effectiveness of the proposed joint phase noise and channel estimation technique. We compare the performance of our proposed technique to existing methods, namely BEM and Spline, as well as the case where only stage 1 of the proposed technique is used. We compare the techniques in terms of bit error rate (BER), error vector magnitude (EVM) and normalized mean-squared error (NMSE) of the channel estimates. Monte Carlo simulation is used to average the results over $10^5$ random channel instances for each simulation. 

A carrier frequency of $f_\mathrm{c} = 5.9$ GHz, a transmission bandwidth of 7.68~MHz and a delay-Doppler grid size of $M=128$ and $N=32$ are considered. The 5G NR Tapped Delay Line C (TDL-C) channel model with a delay spread of 100~ns \cite{3gpp_TS38901} is used. We consider two scenarios for the maximum Doppler shifts, a static scenario where the velocity is set to 0~km/h and a high mobility scenario where the relative velocity between the transmitter and the receiver is set to 500~km/h. The Doppler shifts are generated using Jakes' model \cite{jakes_model}. For the phase noise, unless otherwise stated, we consider a PLL with $F_{\rm{PLL}} = 1\times10^6$ and $\beta_{\rm{PN}} = 2\times10^3$. At the receiver, we use two equalization methods, linear equalization and non-linear equalization with successive interference cancellation (SIC). For linear equalization, we use a simple MMSE equalizer \cite{OTFS_MMSE}. For non-linear equalization, we use the state-of-the-art least-squares minimum residual with interference cancellation (LSMR-IC) algorithm proposed by the authors of \cite{OTFS_LSMR}. When using LSMR-IC, we set the parameters $I_{\rm{ic}} = 10$ and $I_{\rm{lsmr}} = 20$, which correspond to the number interference cancellation iterations and LSMR iterations, respectively. For the BEM method, we choose the number of basis functions based on the maximum Doppler spread and the phase noise bandwidth, i.e. $Q_{\rm{BEM}}=\lceil 2k_\mathrm{over}M_{\rm T}N(f_{\rm D}+\beta_{\rm{pn}})T_{\rm s}\rceil+1$, where $k_\mathrm{over}\geq 1$ is the BEM oversampling factor.

\begin{figure}[t]
    \centering
    \includegraphics[width=0.75\linewidth]{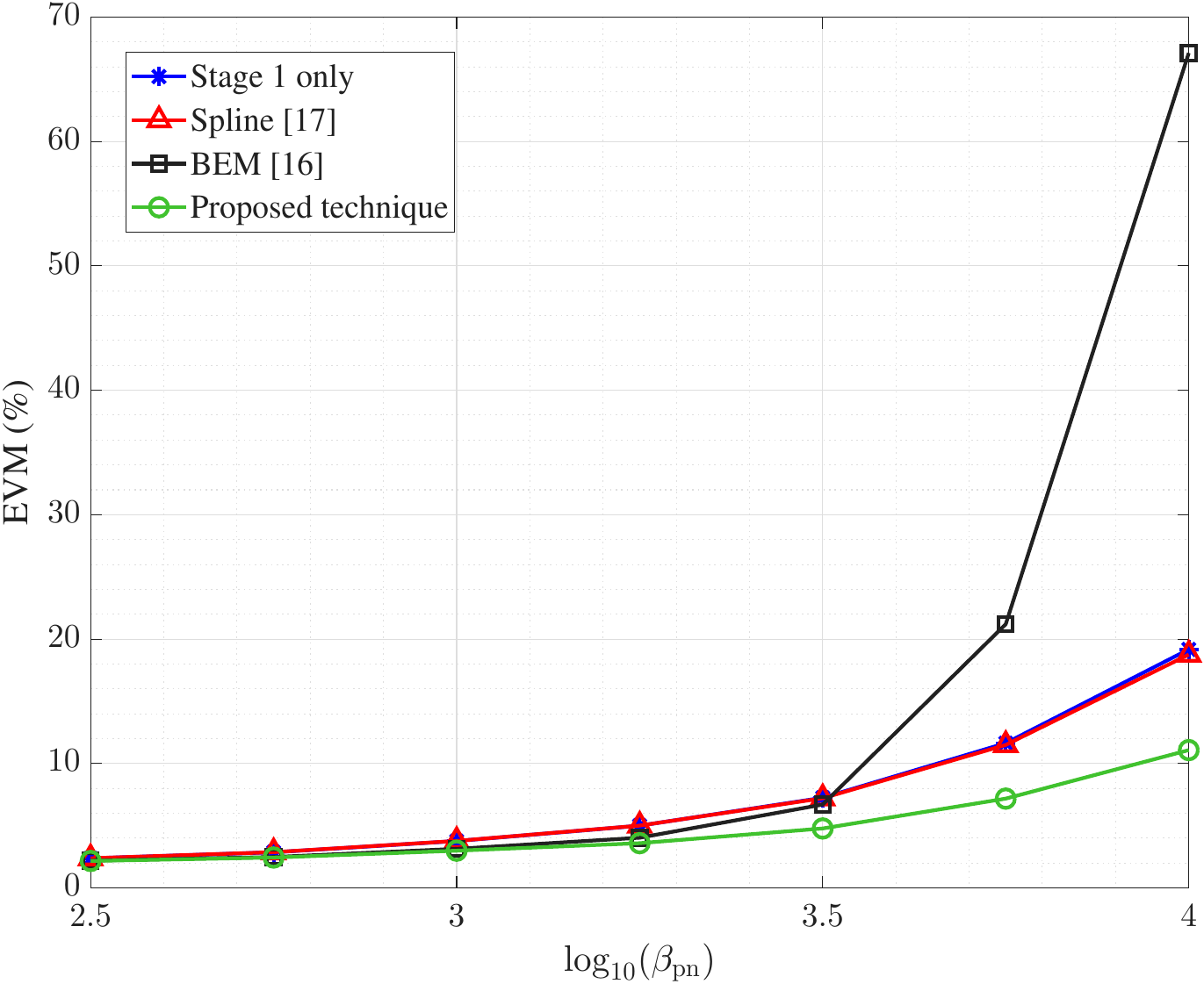}
    \caption{Comparison of the EVM performance of the proposed technique, BEM \cite{Sanoop_PCP}, Spline \cite{Thaj_OTSM_2021} and  Stage 1 only for different phase noise levels with no Doppler spread.}
    \label{fig:SINR_PN_only}
\end{figure}

\begin{figure}[t]
    \centering
    \includegraphics[width=0.75\linewidth]{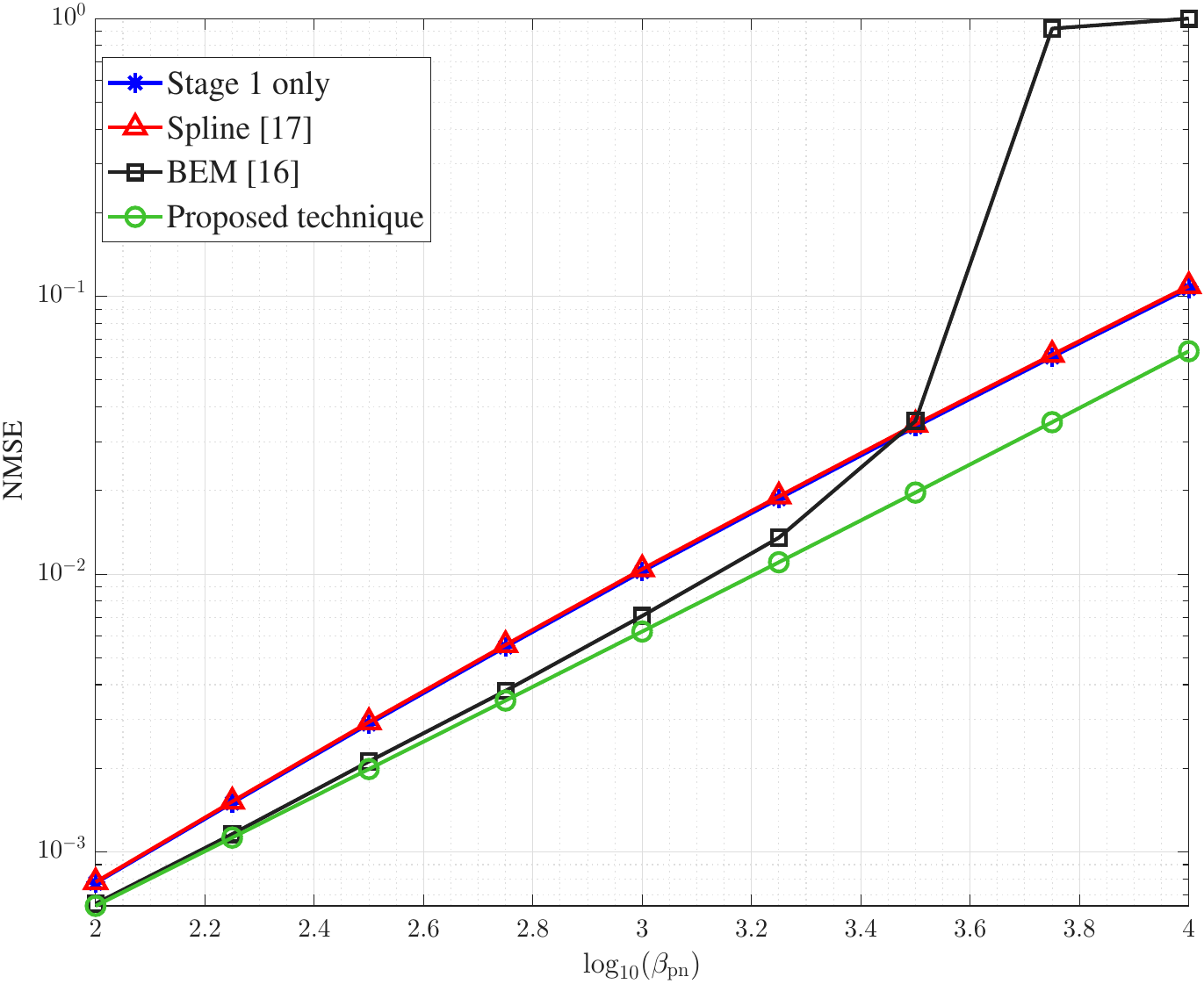}
    \caption{Comparison of the NMSE performance of the proposed technique, BEM \cite{Sanoop_PCP}, Spline \cite{Thaj_OTSM_2021} and  Stage 1 only for different phase noise levels with no Doppler spread.}
    \label{fig:NMSE_PN_only}
\end{figure}

\begin{figure}[t]
    \centering
    \includegraphics[width=0.75\linewidth]{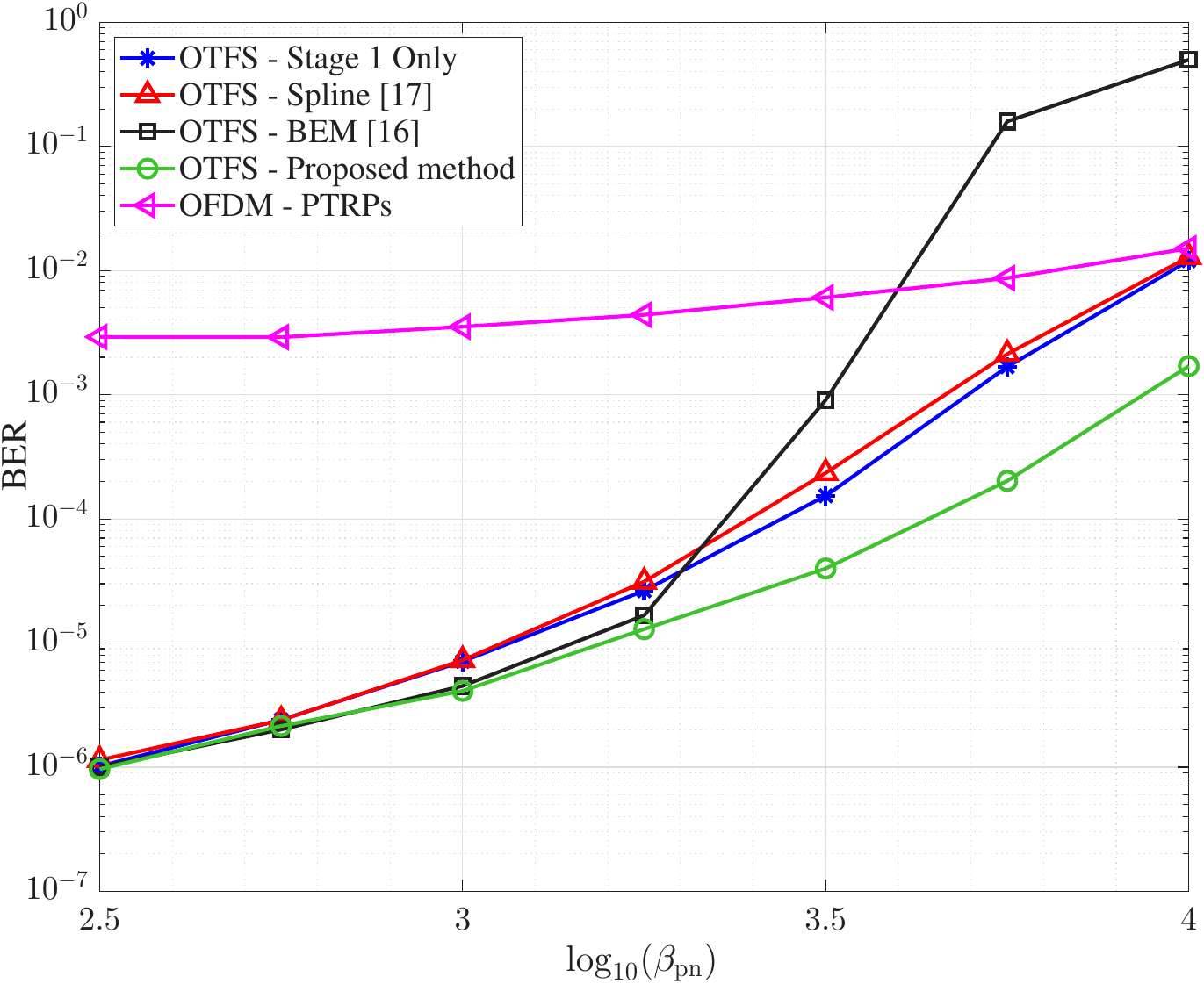}
    \caption{Comparison of the BER performance of the proposed technique, BEM \cite{Sanoop_PCP}, Spline \cite{Thaj_OTSM_2021} Stage 1 only, and OFDM using PTRPs for different phase noise levels with no Doppler spread where 4-QAM is used.}
    \label{fig:BER_OTFS_OFDM}
\end{figure}

\begin{figure}[t]
    \centering
    \includegraphics[width=0.75\linewidth]{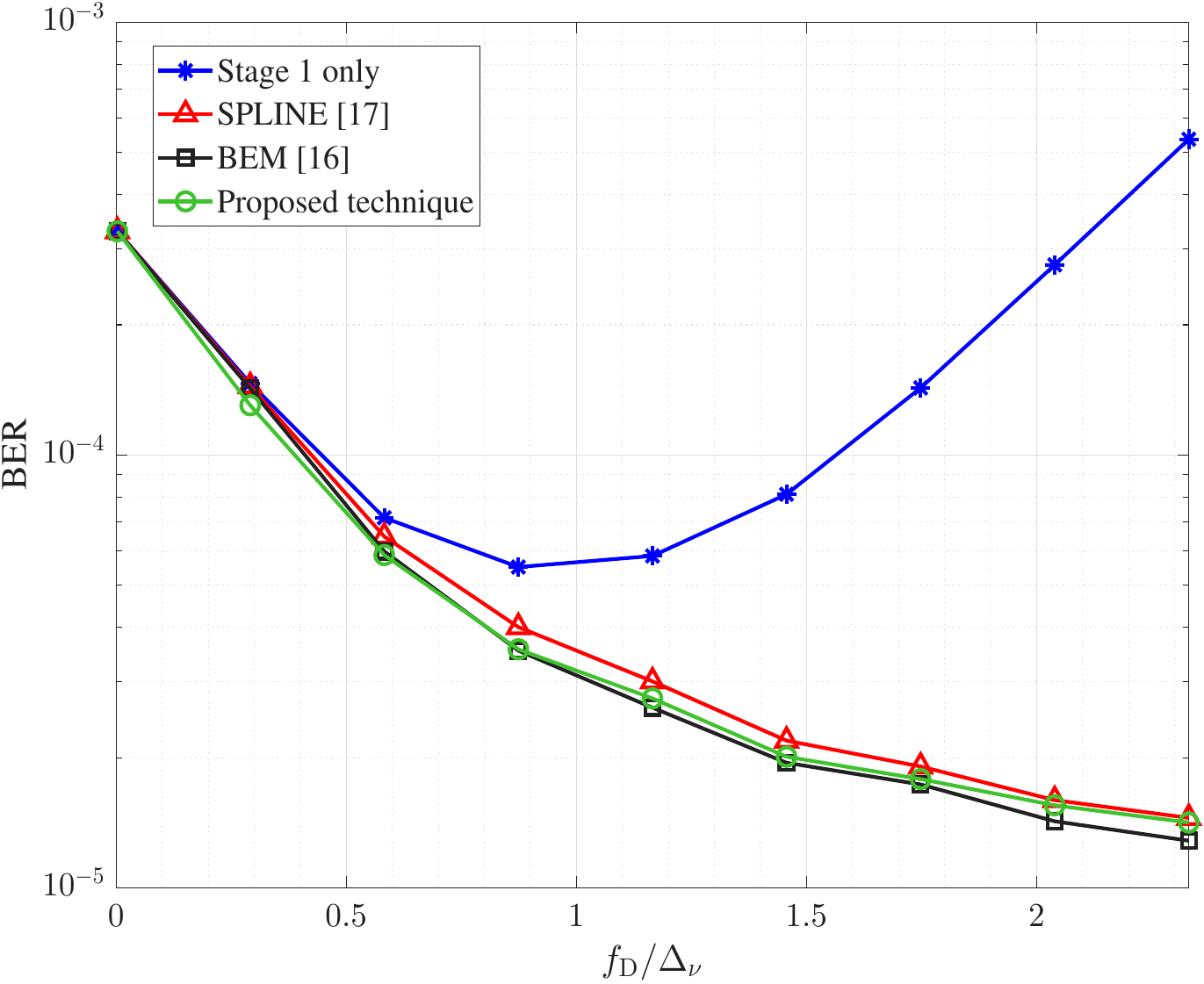}
    \caption{Comparison of the BER performance of the proposed technique, BEM \cite{Sanoop_PCP}, Spline \cite{Thaj_OTSM_2021} and Stage 1 only for different maximum Doppler shifts and no phase noise present 4-QAM is used.}
    \label{fig:BER_dopp}
\end{figure}

In order to focus on the impact of phase noise only, the results shown in Fig. \ref{fig:SINR_PN_only} and Fig. \ref{fig:NMSE_PN_only} concern the scenario where there is no Doppler spread, i.e., $f_{\rm D} = 0$. Fig. \ref{fig:SINR_PN_only} shows the EVM of the received symbols after demodulation and equalization as the phase noise bandwidth increases. For this analysis, we consider a fixed SNR level of 20~dB and a receiver using MMSE equalization. It can be seen that our proposed method outperforms the existing methods, especially at higher phase noise levels.  Spline provides negligible improvement over the partial estimation case, indicating that it may be inappropriate for estimating phase noise. An interesting observation is that BEM provides similar performance to proposed method at low phase noise levels but as phase noise increases, BEM faces severe performance degradation. This is because at high phase noise levels, sample-to-sample channel variations reaches a level that BEM cannot capture. Therefore, BEM is ill-suited for estimating phase noise. Fig. \ref{fig:NMSE_PN_only} shows the NMSE of the estimated channel as the phase noise bandwidth increases. As can be seen, our proposed method provides the best performance compared to the other methods. Once again, it can be seen that Spline provides no gain over the case where only stage 1 has been used and that the BEM performance degrades significantly as phase noise increases. This further corroborates the finding from Fig. \ref{fig:SINR_PN_only} that BEM and Spline are not suitable for estimating phase noise in OTFS.

\begin{figure}[t]
    \centering
    \includegraphics[width=0.75\linewidth]{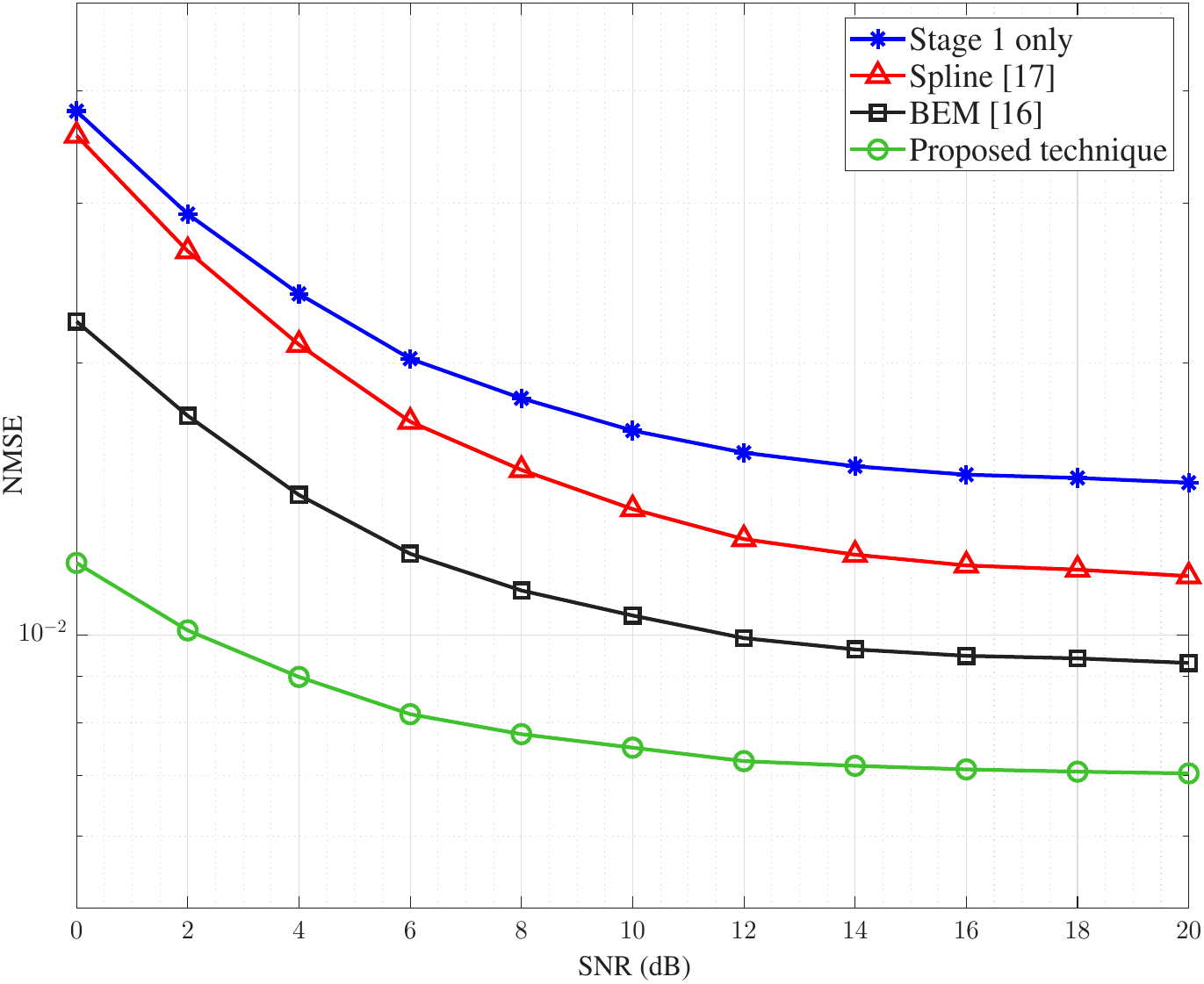}
    \caption{Comparison of the NMSE performance of the proposed technique, BEM \cite{Sanoop_PCP}, Spline  \cite{Thaj_OTSM_2021} and Stage 1 only, for different SNR levels at at $\beta_{\rm{pn}} = 1\times10^3$.}
    \label{fig:NMSE_SNR}
\end{figure}

\begin{figure}[t]
    \centering
    \includegraphics[width=0.75\linewidth]{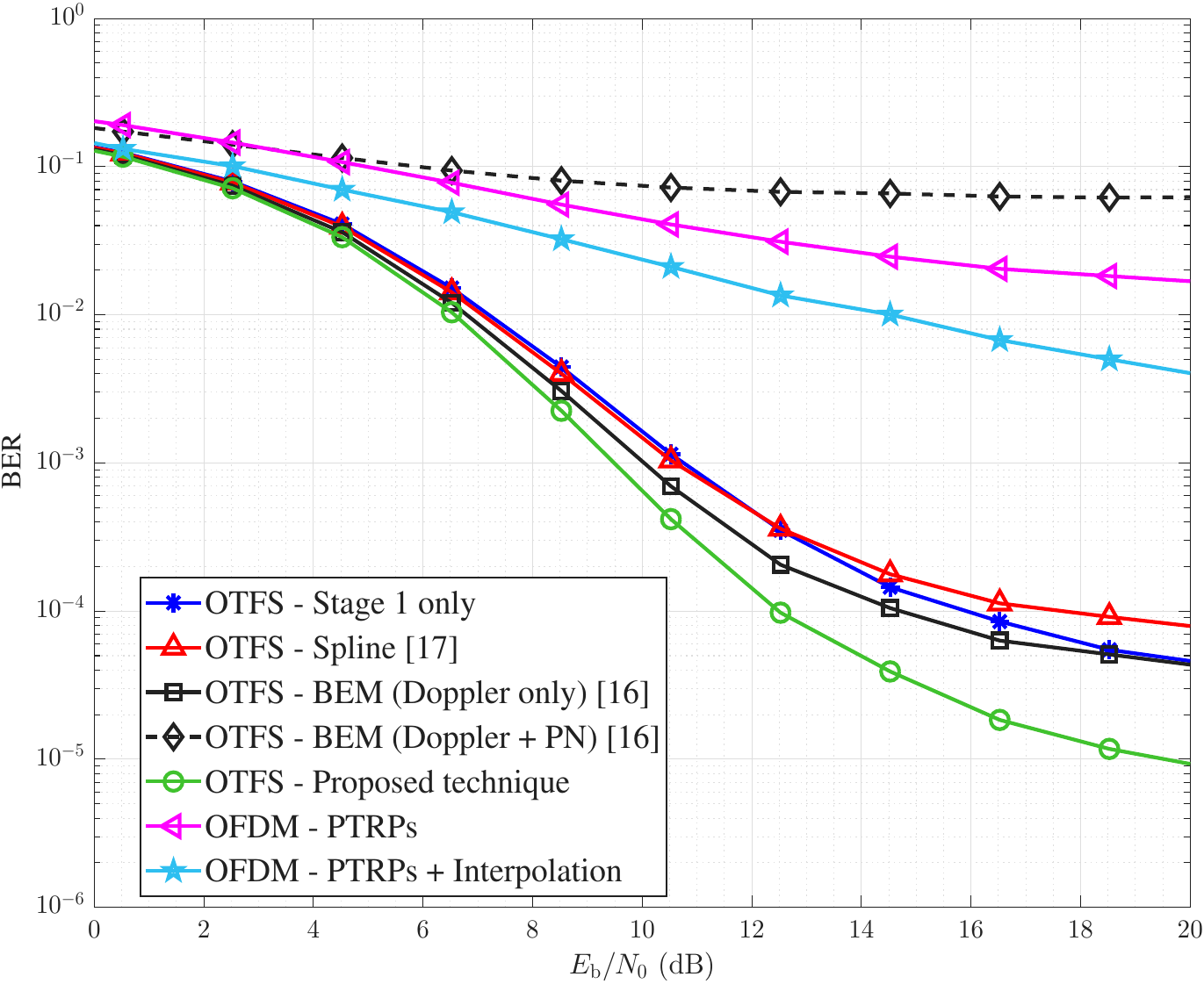}
    \caption{Comparison of the BER performance of the proposed technique, BEM \cite{Sanoop_PCP}, Spline \cite{Thaj_OTSM_2021}, Stage 1 only and OFDM using PTRPs with different SNR levels and $\beta_{\rm{pn}} = 2\times10^3$, for the case where a 4-QAM is used.}
    \label{fig:BER_4qam}
\end{figure}

\begin{figure}[t]
    \centering
    \includegraphics[width=0.75\linewidth]{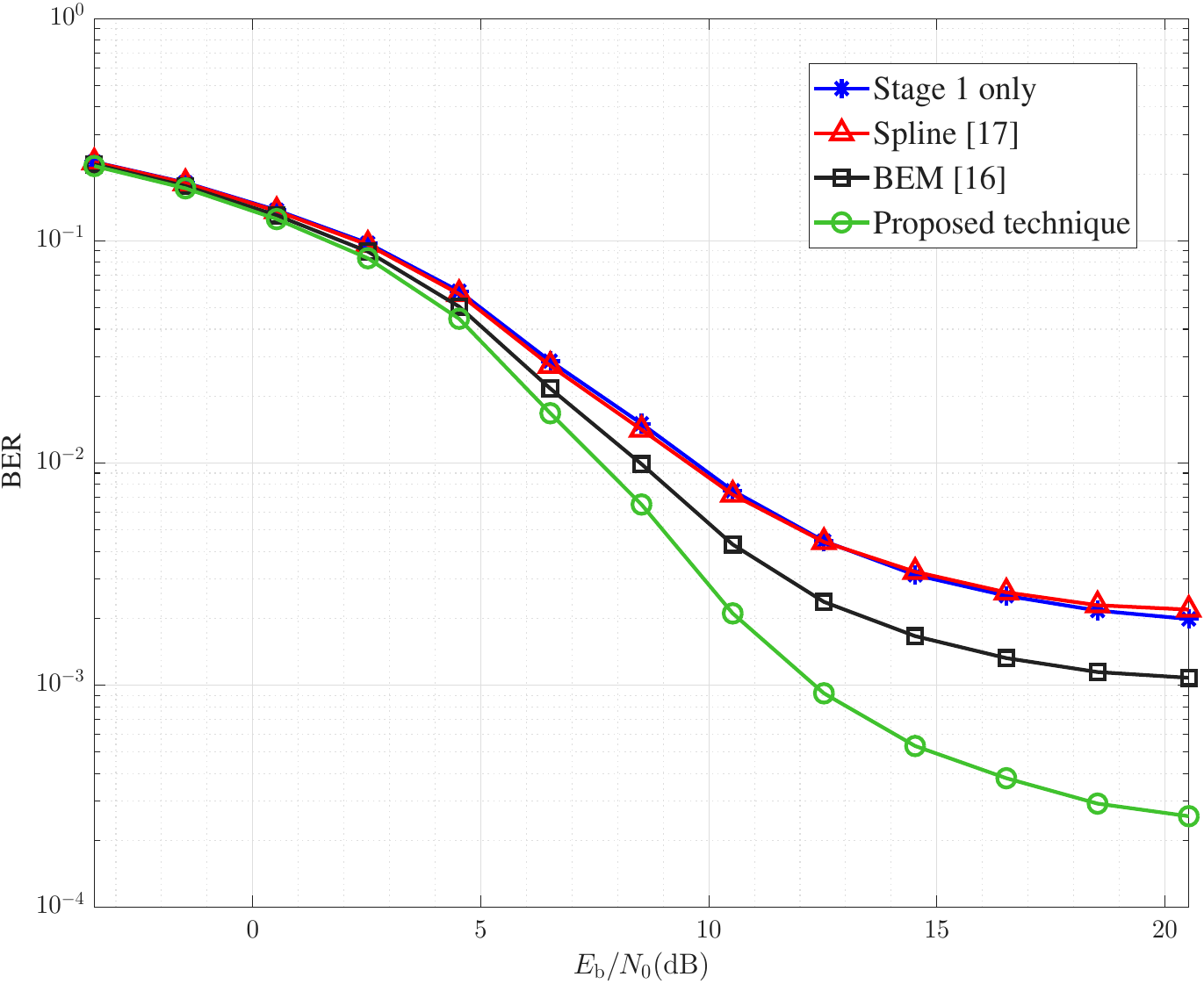}
    \caption{Comparison of the BER performance of the proposed technique, BEM  \cite{Sanoop_PCP}, Spline  \cite{Thaj_OTSM_2021} and Stage 1 only, with different SNR levels and $\beta_{\rm{pn}} = 5\times10^3$, for the case where a 4-QAM is used.}
    \label{fig:BER_4qam_5k}
\end{figure}

\begin{figure}[t]
    \centering
    \includegraphics[width=0.75\linewidth]{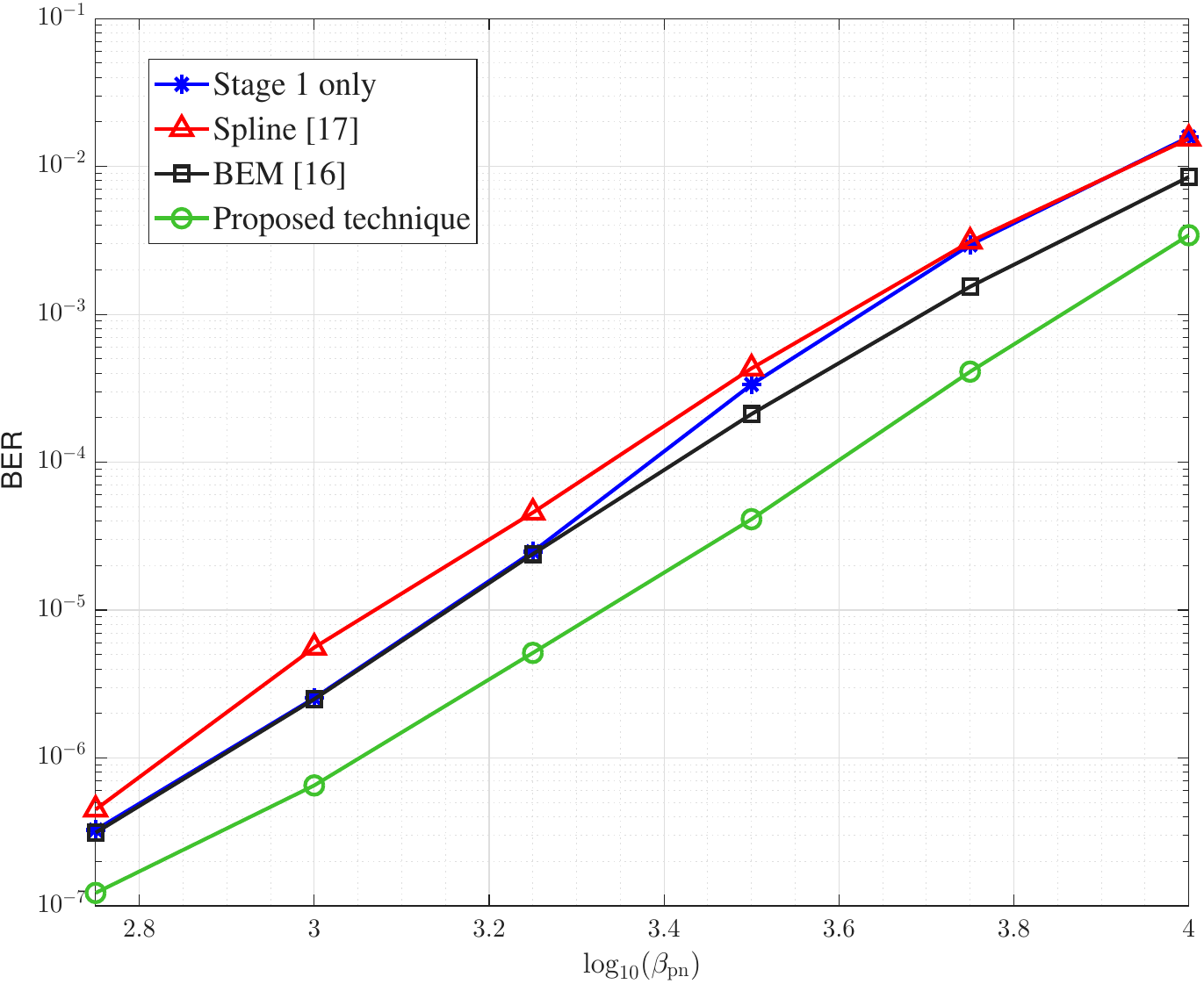}
    \caption{Comparison of the BER performance of the proposed technique, BEM \cite{Sanoop_PCP}, Spline \cite{Thaj_OTSM_2021} and Stage 1 only, with different phase noise levels for the case where a 4-QAM is used.}
    \label{fig:BER_PN}
\end{figure}

Fig. \ref{fig:BER_OTFS_OFDM} shows the BER performance of OTFS as well as OFDM using phase tracking reference pilots as a function of $\beta_{\rm{PN}}$ with a fixed SNR of 24~dB and no Doppler spread. This figure shows that OTFS, using all the methods under consideration, generally outperforms OFDM with PTRPs until phase noise is extremely high. At extremely high levels of phase noise, the Stage 1 only and Spline approaches achieve the same performance as OFDM with PTRPs, while the proposed method still provides an order of magnitude performance gains over them. It can be seen that increasing phase noise has a much greater relative effect on the performance of OTFS, where OTFS loses a significant portion of it's performance gains over OFDM. This further demonstrates that OTFS is in fact quite sensitive to phase noise. These also demonstrate the added benefits of the proposed method, even at extremely high levels of phase noise over the existing OTFS methods in the literature.

Next, we investigate the performance of our proposed method in the case where phase noise is not present , i.e, $\beta_{\rm{pn}} = 0$, and there is only Doppler spread due to mobility. Fig. \ref{fig:BER_dopp} shows the performance of our proposed technique compared to the benchmarks in terms of BER as maximum Doppler spread, normalized to the Doppler spacing $\Delta{\nu}$, increases at an SNR of 20~dB. It can be seen that as the Doppler spread increases, the performance of partial estimation method degrades significantly and that each of the more advanced methods provides improved performance. The proposed technique methods provide approximately the same performance as both BEM and Spline. It is important to note that for each of these three techniques, BER performance improves as Doppler spread increases due to increased channel diversity \cite{OTFS_diversity}. 

\begin{figure}[t]
    \centering
    \includegraphics[width=0.75\linewidth]{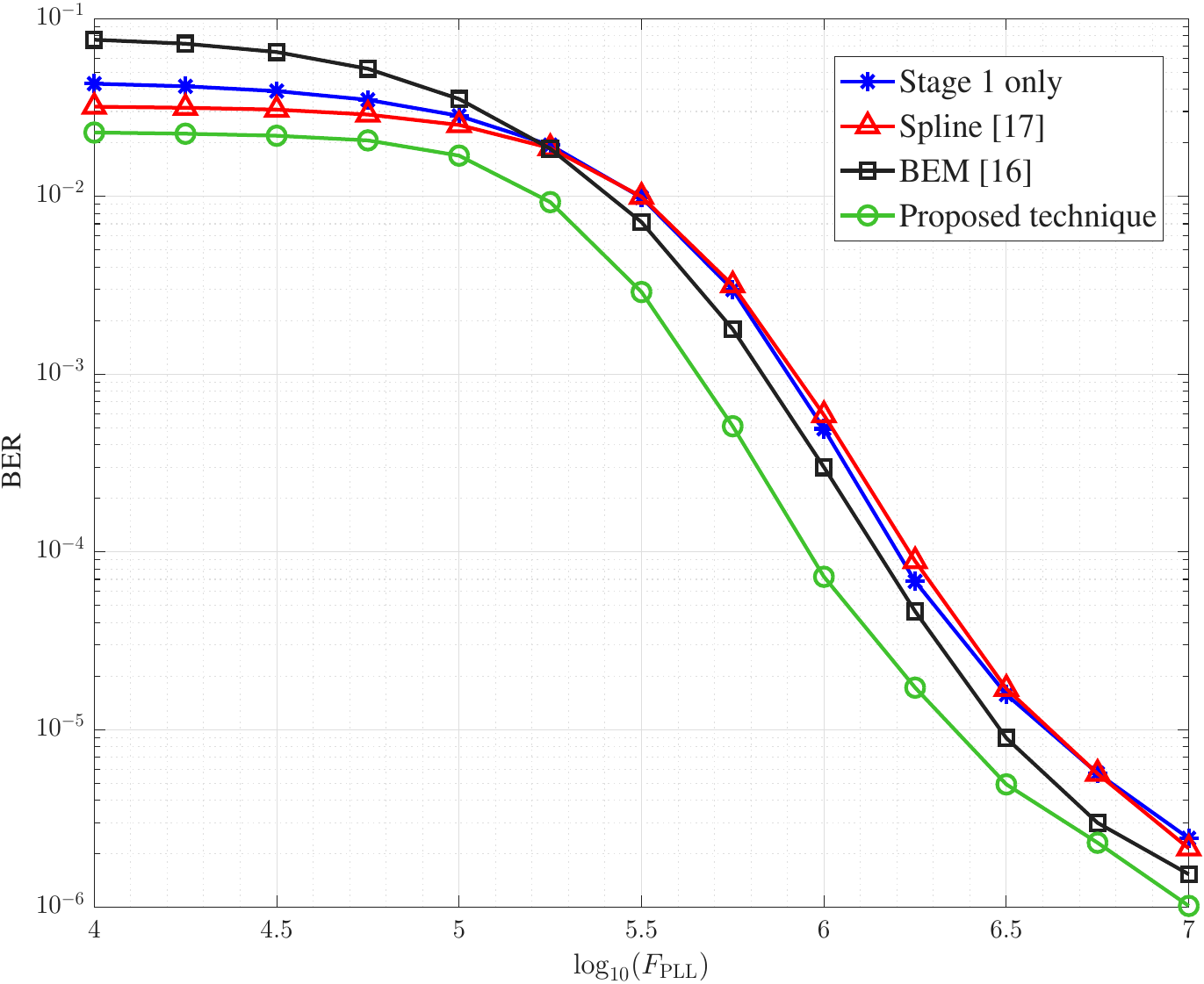}
    \caption{Comparison of the BER performance of the proposed technique, BEM \cite{Sanoop_PCP}, Spline \cite{Thaj_OTSM_2021} and Stage 1 only, with different levels of PLL filtering.}
    \label{fig:BER_pll}
\end{figure}

\begin{figure}[t]
    \centering
    \includegraphics[width=0.75\linewidth]{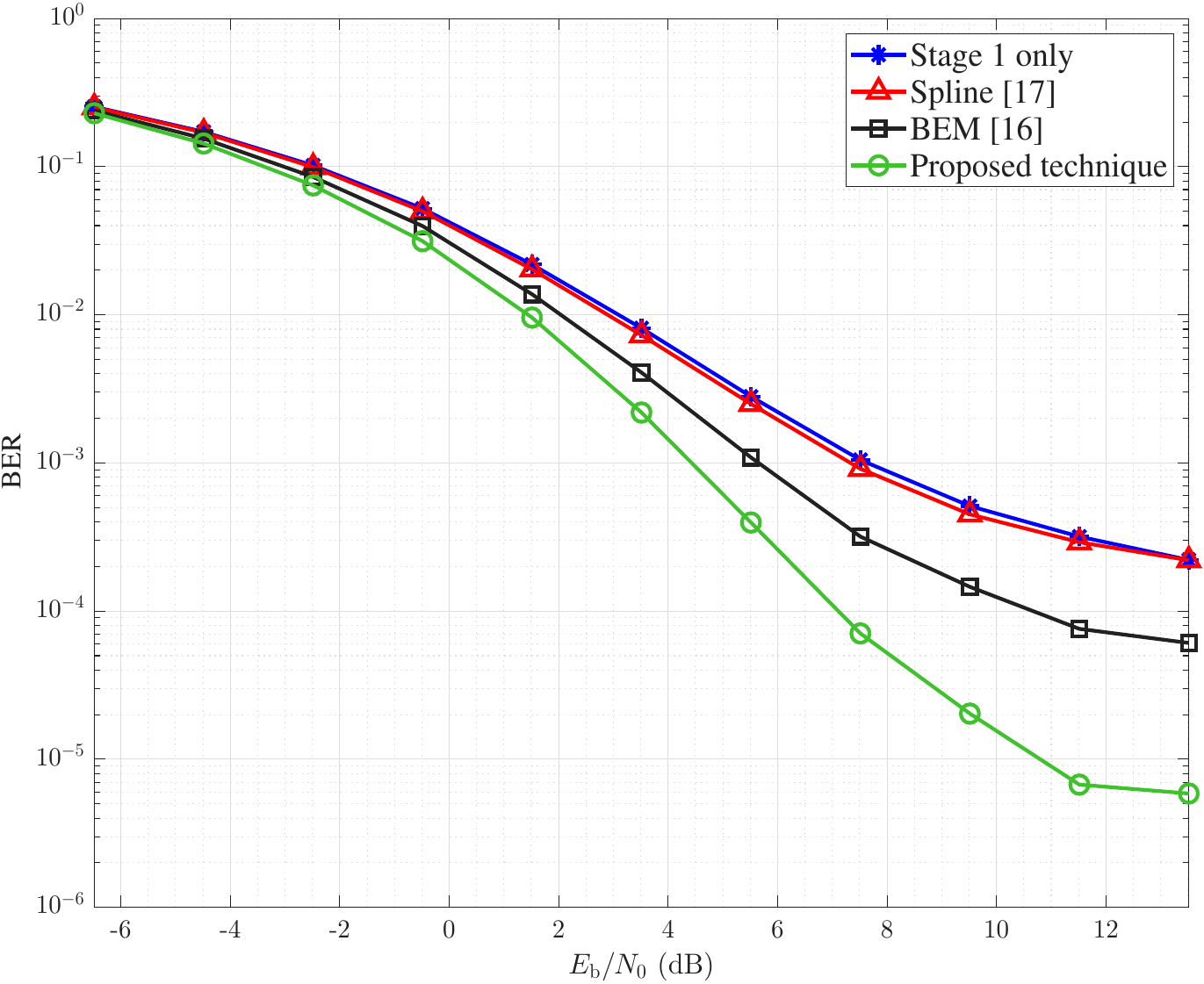}
    \caption{Comparison of the coded BER performance of the proposed technique, BEM \cite{Sanoop_PCP}, Spline \cite{Thaj_OTSM_2021} and Stage 1 only, with different SNR levels and $\beta_{\rm{pn}} = 1\times10^4$, for the case where a 4-QAM is used.}
    \label{fig:4_qam_coded}
\end{figure}

\begin{figure}[t]
    \centering
    \includegraphics[width=0.75\linewidth]{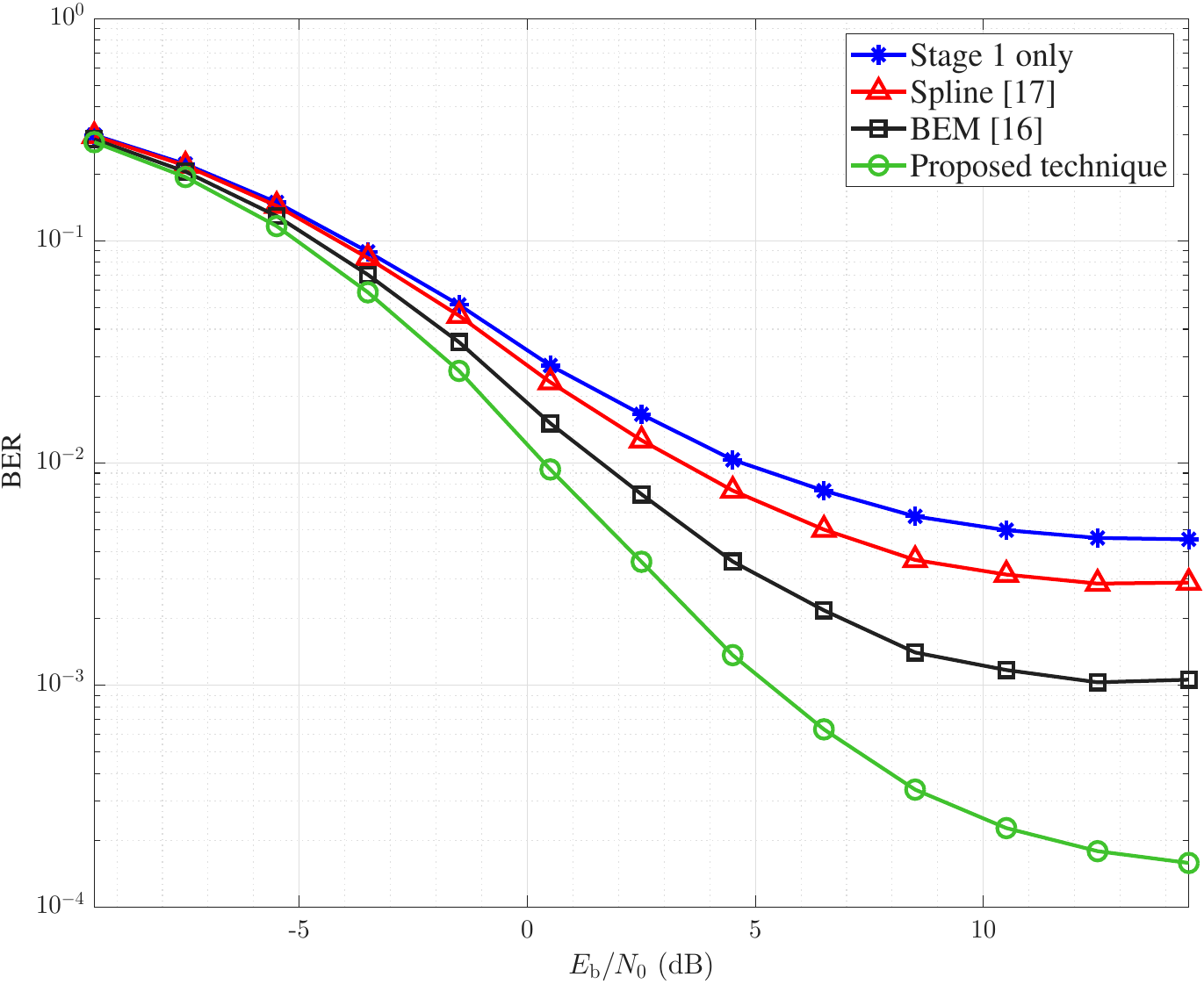}
    \caption{Comparison of the BER performance of the proposed technique, BEM \cite{Sanoop_PCP}, Spline \cite{Thaj_OTSM_2021} and Stage 1 only, with different SNR levels and $\beta_{\rm{pn}} = 5\times10^3$, for the case where a 16-QAM is used.}
    \label{fig:16-qam_coded}
\end{figure}

The results shown in Fig. \ref{fig:BER_4qam} to Fig. \ref{fig:16-qam_coded} concern the scenario where both phase noise and Doppler spread are present. We consider a relative velocity between the transmitter and receiver of 500~km/h in these simulations. Since the previous results in Fig. \ref{fig:SINR_PN_only} show that considering the phase noise bandwidth when forming the BEM coefficients leads to poor performance, we now consider only the maximum Doppler spread for choosing the number of BEM basis functions. Thus, we form the BEM coefficients as $Q = \lceil 2MN T_{\rm{s}} k_{\rm{over}} f_{\rm{D}} + 1\rceil .$
Additionally, to improve detection performance for all the considered techniques, the results shown in the rest of this section stem from simulations where the non-linear equalization method LSMR-IC is used.

Fig. \ref{fig:NMSE_SNR} shows the NMSE of the channel estimates for our proposed method compared to the benchmarks. For these results a phase noise bandwidth of 1~kHz is used. It can be seen that our proposed technique outperforms the benchmark methods and achieves a 10~dB gain over the BEM method at an NMSE of $10^{-2}$. Next we compare the BER performance of OTFS to OFDM when there is both Doppler and phase noise present.  For this result, we also consider an approach where we use BEM to interpolate between time domain channel estimates to obtain a full channel estimate for the entire OFDM frame, as was done in \cite{Huang_OTFS_OFDM}. This approach is referred to in the figure as ``OFDM - PTRP + Interpolation". Fig. \ref{fig:BER_4qam} shows BER for an uncoded system where 4-QAM modulation was used with a fixed $\beta_{\rm{PN}} = 2\times10^3$ and a velocity of 500~km/h. This figure shows that OTFS provides order of magnitude performance gain over OFDM with PTRP, both with and without interpolation, in the presence of both Doppler and phase noise. Additionally, it can be seen from Fig. \ref{fig:BER_4qam} that our proposed method significantly outperforms the competing schemes, providing 2~dB of gain over the BEM method at a BER of $1\times 10^{-4}$ and up to 5~dB as the BEM reaches the error floor at BER of $4\times 10^{-5}$. 
 Fig. \ref{fig:BER_4qam_5k} show the BER results for an uncoded system where 4-QAM modulation was used and the phase noise bandwidth is set equal to 5~kHz. It can be seen that at a higher level of phase noise, the performance across all of the methods degrades. However, our proposed method still provides significantly better performance with approximately 8~dB of gain compared to the BEM method at a BER of $1\times 10^{-3}$. 

Fig. \ref{fig:BER_PN} shows the BER results for an uncoded system where 4-QAM modulation is used as phase noise bandwidth increases. For this analysis, we consider an SNR of 24~dB and a relative velocity of 500 km/h between the transmit and receive antennas. It can be seen that the BER performance of all of the methods degrades as phase noise increases. This is in contrast to Doppler spread, which leads to improved BER performance as it increases. This is because phase noise does not lead to increased channel diversity due to it's single tap nature. However, despite the degradation in performance across all of the methods, we can also see that our proposed method provides the best performance of all the techniques shown. For example, at a BER of $1\times 10^{-3}$, our proposed method can tolerate a phase noise with a bandwidth which is 1.5~kHz wider than the amount that the BEM method can handle. Additionally, while BEM provides negligible performance gains at low levels of phase noise, i.e. $\beta_{\rm{pn}}$ of 1~kHz, our proposed method still provides a superior performance, demonstrating that it has utility in both high and low phase noise scenarios.

Fig. \ref{fig:BER_pll} shows the performance of our proposed method compared to the benchmarks for different PLL filter designs. In this analysis, we consider fixed phase noise bandwidth of 3~kHz, SNR of 20~dB and a relative velocity of 500~km/h between transmitter and receiver. At lower values of $F_{\rm{PLL}}$, the PLL does not filter the phase noise well and the system acts like a  free-running oscillator. As $F_{\rm{PLL}}$ increases, the PLL is suppressing more of the VCO phase noise and the system becomes closer to an AWGN process in behavior. It can be seen from the figure that our proposed method provided superior performance in all scenarios. It should be noted that when $F_{\rm{PLL}}$ is large the performance gains are quite small. This result is due to the assumption of a noiseless reference oscillator and noisy VCO oscillator in the PLL. When phase noise is present at the VCO, it experiences a high pass filter response and a large $F_{\rm{PLL}}$ value suppresses it well. However, in practical systems, there is also some phase noise present at the reference oscillator. This noise would experience a low-pass filter response and thus, a large $F_{\rm{PLL}}$ would not suppress it. In practical systems, the value of $F_{\rm{PLL}}$ is chosen to strike a balance between suppression of VCO phase noise and reference oscillator phase noise. Therefore, we conclude that our proposed method is still necessary for practical values of $F_{\rm{PLL}}$.

Fig. \ref{fig:4_qam_coded} and Fig. \ref{fig:16-qam_coded} show the BER results for the coded system where 4-QAM modulation and 16-QAM modulation are used respectively. In this analysis, we utilize of a 1/2 rate convolutional encoder to improve BER performance. This use of coding allows us to investigate performance in more severe phase noise environments. In Fig. \ref{fig:4_qam_coded} the phase noise bandwidth is set equal to 10~kHz and in Fig. \ref{fig:16-qam_coded} it is set equal to 5~kHz. It can be seen from Fig. \ref{fig:4_qam_coded} that while the use of coding improves performance of all the methods, our proposed method still outperforms the competing schemes. Our proposed technique provides 4~dB of gain over the BEM method at a BER of $1\times 10^{-4}$ and an order of magnitude improvement in detection performance at an $E_{\rm{b}}/N_{\rm{0}}$ of 11~dB. In Fig. \ref{fig:16-qam_coded}, it can be seen that at a higher modulation order our proposed method still provides significant gains over the other methods under study. Our proposed technique provides approximately 6~dB of gain over the BEM method as the BEM reaches an error floor at a BER of $1\times 10^{-3}$. These results demonstrate that in coded systems, our proposed technique provides significant gains over existing methods in the literature.

\section{Conclusion}
In this paper, we investigated the effect of phase noise on OTFS systems. We analyzed the interference cause by phase noise in the delay-Doppler domain and demonstrated that OTFS is sensitive to phase noise without appropriate compensation. To do this we derived expressions for the interference power and SINR due to phase noise in the delay-Doppler domain. These expressions are presented for three phase noise scenarios; the free-running oscillator, the continuous-time PLL and the discrete-time PLL. Additionally, we derived a closed-form expression for the interference power for a system using a free-running oscillator. We then propose a novel technique for joint phase noise and channel estimation using a Wiener filtering based approach. This method is a 2 stage process where in stage 1 a partial channel estimate is obtained via a single impulse pilot in the delay-Doppler domain. The second stage then uses Wiener filtering, based on the statistical nature of the Doppler spread channel and the oscillator phase noise, to estimate the full effective channel. We have presented a range of numerical results which demonstrate the superiority of our proposed technique over existing methods in the literature in terms of BER, NMSE and EVM. Our numerical results demonstrate the superiority of our proposed technique in both high-mobility and low-mobility scenarios, as well as both coded and uncoded systems.

\bibliographystyle{IEEEtran}
\bibliography{biblio}

\end{document}